\documentclass[Universe,article,submit,moreauthors, pdftex,10pt,a4paper]{mdpi} 
\firstpage{1} 
\articlenumber{x}
\doinum{10.3390/------}
\pubvolume{xx}
\pubyear{2016}
\copyrightyear{2016}
\history{Received: date; Accepted: date; Published: date}
\usepackage{amsfonts}
\usepackage{tensor}
\usepackage{mathrsfs}
\usepackage{slashed}
\usepackage{amssymb}
\usepackage[normalem]{ulem}
 \theoremstyle{mdpi}
 \newcounter{thm}
 \newcounter{ex}
 \newcounter{re}
 \theoremstyle{mdpidefinition}

\def\a{\alpha}
\def\b{\beta}
\def\g{\gamma}

\def\d{\delta}

\def\e{\epsilon}

\def\l{\lambda}

\def\m{\mu}
\def\n{\nu}
\def\x{\xi}
\def\X{\Xi}

\def\r{\rho}

\def\vf{\varphi}

\def\cF{{\cal F}}

\def\cK{{\cal K}}
\def\cL{{\cal L}}
\def\cM{{\cal M}}
\def\cN{{\cal N}}
\def\cO{{\cal O}}
\def\cP{{\cal P}}
\def\cQ{{\cal Q}}
\def\cR{{\cal R}}

\def\cT{{\cal T}}

\def\cW{{\cal W}}

\def\be{\begin{equation}}
\def\ee{\end{equation}}
\def\bea{\begin{eqnarray}}
\def\eea{\end{eqnarray}}
\def\beal{\begin{equation}\begin{aligned}}
\def\eeal{\end{aligned}\end{equation}}
\def\ba{\begin{array}}
\def\ea{\end{array}}
\def\bec{\begin{center}}
\def\ec{\end{center}}
\def\ba{\begin{align}}
\def\ena{\end{align}}
\def\ft{\footnote}

\def\Ddot{D\cdot}

\def\nn{\nonumber}
\def\pe{\prime}

\def\12{\frac{1}{2}}
\def\fr{\frac}
\def\pr{\partial}

\newcommand{\bin}[2]{{#1 \choose #2}}

\Title{Asymptotic Charges at Null Infinity in Any~Dimension}

\Author{Andrea Campoleoni $^{1}$\orcidA{}, Dario Francia $^{2}$\orcidB{} and Carlo Heissenberg $^{2}$\orcidC{}} 
\AuthorNames{Andrea Campoleoni, Dario Francia and Carlo Heissenberg}

\address{%
$^{1}$ \quad Institut f\"ur Theoretische Physik, ETH Zurich, Wolfgang-Pauli-Strasse, 27 8093 Z\"urich, Switzerland; campoleoni@itp.phys.ethz.ch\\
$^{2}$ \quad Scuola Normale Superiore and INFN, Piazza dei Cavalieri, 7 I-56126 Pisa, Italy; \hspace{40pt}dario.francia@sns.it, carlo.heissenberg@sns.it}
\abstract{We analyse the conservation laws associated with large gauge transformations of massless fields in Minkowski space. Our aim is to highlight the interplay between boundary conditions and finiteness of the asymptotically conserved charges in any space-time dimension, both even and odd, greater than or equal to three. After discussing nonlinear Yang--Mills theory and revisiting linearised gravity, our investigation extends to cover the infrared behaviour of bosonic massless quanta of any~spin.
}

\begin{document}

\setcounter{tocdepth}{1}
\tableofcontents
 
\vspace{15pt}
%%%%%%%%%%%%%%%%%%%%%%%%%%%%%%%%%%%%%%%%%%%%%%%%%%%%%%%%%%%%%%%%%%%%%
\section{Introduction}\label{sec:intro}
%%%%%%%%%%%%%%%%%%%%%%%%%%%%%%%%%%%%%%%%%%%%%%%%%%%%%%%%%%%%%%%%%%%%%
 
  In a previous work \cite{super}, we investigated the asymptotic symmetries of massless bosons of spin greater than two in four-dimensional Minkowski spacetime. We found that, upon assigning suitable boundary conditions on the rank$-s$ symmetric tensors $\vf_{\m_1 \, \cdots \, \m_s}$, the asymptotic Killing equations for the rank$-(s-1)$ gauge parameters $\xi_{\m_1 \, \cdots \, \m_{s-1}}$ admit an infinite-dimensional set of solutions providing counterparts of the supertranslations  emerging for spin-two fields in asymptotically flat spaces \cite{BMS, Sachs_Symmetries}. In particular, in strict analogy with the spin-two case \cite{Strominger_Invariance,Weinberg-BMS}\footnote{See also \cite{Strominger_rev} for a general review and more references.}, we found that Weinberg's soft theorem for any spin \cite{Weinberg_64,Weinberg_65} could be derived as a consequence of the higher-spin supertranslation Ward identities. In addition, we studied the full asymptotic Killing tensor equation in any space-time dimension $D$ for spin-three fields, finding in particular proper counterparts of superrotations in four dimensions \cite{Barnich_Revisited, Barnich_BMS/CFT}. 
  
In the present work, our goal is twofold: (1) to extend the analysis of asymptotic symmetries for all spins to arbitrary values of the space-time dimension; (2) to compute the resulting charges and check their finiteness, thus proving the consistency of our choice of falloffs.

 In order for our treatment to be as homogeneous as possible for any $D$, here we shall not make use of the notion of conformal null infinity. Indeed, its construction was shown to be obstructed in odd-dimensional spacetimes containing radiation because of singularities appearing in the components of the Weyl tensor of the unphysical space \cite{gravity_evenD_1, hollands2004}. Differently, focussing on the falloffs of the solutions to the relevant equations of motion results in an exploration of null infinity that is devoid of such issues and thus allows for the same type of analysis in all dimensions \cite{gravity_anyD}.

 Our general procedure can be summarised as follows:  for all spins, we assume a power-like behaviour for the radial dependence of the field components keeping track of  all possible subleading contributions (with some subtleties for the case of Yang--Mills theory in $D=3$, where logarithmic dependence is also taken into account). In addition, we fix our boundary data through a set of Bondi-like conditions that can be interpreted as resulting from on-shell local gauge fixings. In this framework a difference emerges between even and odd dimensionalities: whenever $D$ is odd and greater than four, in order for both the radiation part and the Coulomb part of the solution to be accounted for, one finds that the expansion in powers of $r$ requires both integer and half-integer exponents to be considered. Differently, only integer powers of $r$ are needed whenever $D$ is even. Moreover, radiation and Coulombic contributions behave as $r^{-(D-2)/2}$ and $r^{3-D}$ respectively, and thus actually coincide in $D=4$, thus justifying a separate analysis for dimensionalities  higher than four on the one hand and lower or equal to four on the other. Leading and subleading falloffs are determined by solving the equations of motion, while their consistency relies on checking that the energy flowing to null infinity per unit of retarded time is indeed finite. Once the falloffs are determined, we proceed to compute the asymptotic symmetries and the corresponding charges, while also checking finiteness of the latter.
 
 Proceeding along these lines, in Section~\ref{sec:spin1} we provide a full analysis of the nonlinear Yang--Mills theory in any dimension, starting from $D=3$.   The investigation of  asymptotic symmetries and related aspects in dimensions other than four (both for spin one and spin two) has been performed in a number of works \cite{Einstein-YM Barnich,Mao_evenD, Maxwell d=3 Barnich, Pate_Memory, Kapec:2015vwa, Campiglia_scalars}. With respect to previous  explorations of the Yang--Mills case in any $D$ \cite{Einstein-YM Barnich,Mao_evenD}, here we also add the explicit computation of the charges, while, for the three-dimensional case already discussed in \cite{Maxwell d=3 Barnich}, we include the contribution of radiation. For the four-dimensional analysis of the spin-one case see also \cite{yang-mills_strominger,Maxwell d=4 Strominger,Campiglia_QED,Mao_em,Mao_note,Strominger "Kac",Strominger Color,Adamo Casali}. 

 In Section~\ref{sec:spin2} we revisit the case of asymptotically flat gravity, for which an analysis in any dimension, both even and odd, can be found in \cite{gravity_anyD}. Our review focusses on the linearised theory, which is useful for us in order to set the stage for the ensuing generalisation to higher spins that we first illustrate in Section~\ref{sec:spin3} for the spin-three case. In particular, we complete the analysis in arbitrary dimension presented in \cite{super} by computing the charges corresponding to the asymptotic symmetries. In~Section~\ref{sec:spin-s} we pursue our exploration of the general spin$-s$ case initiated in \cite{super}. In this respect, besides extending the study of large gauge transformations to higher space-time dimensions, we determine explicitly the proper counterpart of superrotations for any spin in $D=4$. In addition, upon solving the equations of motion, we are led to a proposal for the boundary conditions eventually leading to finite asymptotic charges, which is explicitly tested in examples where we illustrate the on-shell cancellations of otherwise divergent terms. 

 The asymptotic symmetries that result from our analysis for all spins in $D > 4$ correspond to the solution to the global Killing tensor equations and thus do not display the infinite-dimensional enhancement observed in $D \leq 4$. While this result is in agreement with similar conclusions drawn for spin two in previous works \cite{angular-momentum,gravity_evenD_2}, it still leaves a number of questions unanswered, starting from the ultimate origin of Weinberg's soft theorem in $D>4$.
 
 While this work was in preparation, however, Ref. \cite{Pate_Memory} appeared, with an alternative treatment of boundary conditions allowing for infinite-dimensional symmetries for linearised gravity in any even dimension, identified both as the origin of Weinberg's result for $D = 2k$ and as the sources of even-dimensional counterparts of the memory effect. (See also \cite{Kapec:2015vwa,Mao_evenD,Garfinkle:2017fre,Campiglia_scalars} for earlier discussions on the matter.)  
  
 The exploration of asymptotic symmetries for arbitrary-spin massless fields in any $D$, which we started in \cite{super} and in the present work, presents a number of open challenges on which we plan to focus our attention in the future. Among the main ones, it ought to be stressed that our linearised analysis does not allow one to get a concrete grasp on the properties of the putative non-Abelian algebra underlying our findings, crucial to the issue of uncovering the physical meaning of such symmetries. This is relevant in particular in order to assess the role of higher-spin asymptotic symmetries in the high-energy regime of string scattering amplitudes (see e.g.~\cite{Gross:1988ue,Moeller:2005ez,Sagnotti:2010at,Sagnotti:2013bha,Casali:2016atr}). In particular, in the latter respect, although once again of general interest in itself, the investigation on the possible infinite-dimensional enhancement of global asymptotic symmetries for all spins in $D>4$ manifests special relevance.
 
\vspace{6pt}
%%%%%%%%%%%%%%%%%%%%%%%%%%%%%%%%%%%%%%%%%%%%%%%%%%%%%%%%%%%%%%%%%%%%%
\section{Yang--Mills Theory}\label{sec:spin1}
%%%%%%%%%%%%%%%%%%%%%%%%%%%%%%%%%%%%%%%%%%%%%%%%%%%%%%%%%%%%%%%%%%%%%

 In this section, we analyse  the equations of motion for Yang--Mills theory in $D$-dimensional Minkowski spacetime expanding their solutions in powers of $1/r$, thereby identifying the data that contribute to colour charge and to colour or energy flux at null infinity. In particular, we complement the related discussions in \cite{yang-mills_strominger, Maxwell d=4 Strominger, Strominger "Kac", Strominger Color, Maxwell d=3 Barnich, Campiglia_QED, Adamo Casali, Mao_note} by providing a unified treatment of all spacetime dimensions, and that in \cite{Einstein-YM Barnich,Mao_evenD} by checking the finiteness of asymptotic charges in any dimension while also including radiation for $D=3$.

We adopt the retarded Bondi coordinates  $(x^\mu)=(u,r,x^i)$, where $x^i$, for $i=1,2,\ldots,n$, denotes~the $n := D-2$ angular coordinates on the sphere at null infinity. In these coordinates, the Minkowski metric reads 
\be \label{bondi-coord}
ds^2 = - du^2 - 2 du dr + r^2 \gamma_{ij}\, dx^i dx^j\,,
\ee
where $\gamma_{ij}$ is the metric of the Euclidean $n$-sphere. The corresponding (flat) spacetime connection is denoted by $\nabla_{\!\mu}$, whose nonzero Christoffel symbols read
\be \label{christoffel}
\Gamma\indices{^i_{jr}}=r^{-1} \delta\indices{^i_j}\,,\qquad
\Gamma\indices{^u_{ij}}=r\, \gamma_{ij}=-\,\Gamma\indices{^r_{ij}}\,,\qquad
\Gamma\indices{^k_{ij}}=\frac{1}{2}\, \gamma^{kl}\left(\partial_i \gamma_{jl}+\partial_j \gamma_{il}-\partial_l \gamma_{ij} \right) .
\ee

The Yang--Mills connection is denoted by
$
\mathcal A_\mu := \mathcal A_\mu^A T^A\, ,
$ 
where the $T^A$ are the generators of a~compact Lie algebra $\mathfrak g$, whose gauge transformation is  $ \delta_\epsilon\mathcal A_\mu = \nabla_{\!\mu} \epsilon + \left[ \mathcal A_\mu, \epsilon \right] $. The corresponding field strength is given by 
\be
\mathcal F_{\mu\nu} = \nabla_{\!\mu} \mathcal A_\nu - \nabla_{\!\nu} \mathcal A_\mu + \left[ \mathcal A_\mu, \mathcal A_\nu \right] ,
\ee
while the field equations are
\be\label{EEOM}
\mathcal G_\nu := \nabla^\mu \mathcal F_{\mu\nu} + \left[ \mathcal A^\mu, \mathcal F_{\mu\nu}\right]=0\,.
\ee

Furthermore, we enforce the radial gauge
\be \label{bondi1}
\mathcal A_r=0 \,,
\ee
which completely fixes the gauge in the bulk. 

%%%%%%%%%%%%%%%%%%%%%%%%%%%%%%%%%%%%%%%%%%
\subsection{Boundary Conditions}
%%%%%%%%%%%%%%%%%%%%%%%%%%%%%%%%%%%%%%%%%%

For $D > 3$, we consider field configurations  $\mathcal A_\mu$ whose asymptotic null behaviour is captured by an expansion\footnote{In the three-dimensional case ($n=1$), to be discussed in Section~\ref{sec:3-4dim_1}, we shall also consider a logarithmic dependence in $r$. For a discussion of the asymptotic behaviour of Maxwell fields in Einstein spacetimes see \cite{Ortaggio}.} in powers of $1/r$, for $r\to\infty$. More explicitly, we parameterise their leading-order terms as follows:
\be\label{leading}
\mathcal A_u (u, r, x^i)= r^a\, A_u(u,x^i) + \mathcal O(r^{a-1})  \,, \qquad
\mathcal A_i (u, r, x^j) = r^b\, A_i(u,x^j) + \mathcal O(r^{b-1})\,.
\ee

In order to determine the leading falloffs, we begin our analysis by substituting the conditions \eqref{leading} into the $u-$component of the equations of motion, $\mathcal G_u=0$. To leading order: 
\be\label{esp_leading}
\begin{aligned}
	&
	- a\, \partial_u A_u\, r^{a-1} 
	- \partial_u D^i A_i\, r^{b-2} 
	+\left[\Delta A_u + a(a+1) A_u\right] r^{a-2} \\
	&
	-\gamma^{ij}\left[A_i,\partial_u A_j\right] r^{2b-2} 
	+\left(- D^i\left[A_u,A_i\right]+\gamma^{ij}\left[\partial_i A_j, A_u\right]\right) r^{a+b-2}
	+\gamma^{ij}\left[A_i,\left[A_j, A_u\right]\right]r^{a+2b-2}=0\,,
\end{aligned}
\ee
where $D_i$ denotes the covariant derivative on the Euclidean $n$-sphere, while $\Delta := D^i D_i$.

 Let us notice that, while the three linear terms in the first line are in principle independent, we~can combine them in pairs upon imposing either $b=a+1$, or $b=a$. As we shall see, the different types of solutions arising from these two options retain relevant physical meaning. Indeed, the first one corresponds to \emph{radiation}, with the familiar falloff\footnote{The $D$-dimensional wave equation
$
-\partial_t^2 f + r^{-n}\partial_r ( r^n \partial_r f ) + \Delta f=0
$, where $t=u+r$,
admits spherically symmetric solutions whose large-$r$ behaviour is $r^{-n/2}\exp(i k u)$.
}
 $\mathcal A_u\sim r^{-n/2}$ of a spherical wave, which also carries a finite amount of energy per unit time through null infinity. The latter, on the other hand, leads to \emph{Coulomb-type} solutions with the characteristic falloff $\mathcal A_u\sim r^{1-n}$ of the Coulomb potential, hence giving rise to a finite contribution to the colour charge.

Let us now discuss the asymptotic behaviour of colour and energy flux integrals. The definition of conserved charges associated with gauge symmetries is a~subtle issue and we shall provide more details on the colour charge at null infinity in Section~\ref{subsec: Charges}. Denoting the surface element of the $n$-sphere with unit radius by
$
d\Omega_n
$,
the $A-$th component of the colour charge at a given retarded time $u$ is expressed as the following integral over the sphere $S_u$ at a~given value of $u$,
\be\label{color_u}
\mathcal Q^A(u)=\lim_{r\to\infty}\int_{S_u} \mathrm{tr} \left(\mathcal F_{ur}T^A\right) r^n d\Omega_n\,.
\ee
The energy flowing across $S_u$ per unit time, on the other hand, can be cast as\footnote{The Yang--Mills Lagrangian for anti-Hermitian fields is $\cL = \frac{1}{4}\mathrm{tr}(\mathcal F_{\mu\nu} \mathcal F^{\mu\nu})$, while the stress-energy tensor has the form $T_{\mu\nu}=-\,\mathrm{tr}\left(\mathcal F_{\mu\alpha}\mathcal F\indices{_\nu^\alpha}\right)+\frac{1}{4}g_{\mu\nu}\mathrm{tr}\left(\mathcal F_{\alpha\beta}\mathcal F^{\alpha\beta}\right)$. The energy flux across $S_u$ is then given by $-\int_{S_u} T\indices{_u^r} r^n d\Omega_n=\int_{S_u} (T_{uu}-T_{ur}) r^n d\Omega_n\, $ as $r\to\infty$.}
\be\label{power_u}
\mathcal P(u)=\lim_{r\to\infty}\int_{S_u} \gamma^{ij}\, \mathrm{tr}\left(\mathcal F_{ui}(\mathcal F_{rj}-\mathcal F_{uj})\right) r^{n-2} d\Omega_n\,.
\ee
The request that this quantity be finite imposes that the fields must go to
zero at infinity in order to compensate for the factor of $r^{\, n-2}$, namely 
\be\label{<0}
a<0\,,\qquad b<0\,,
\ee
whenever $D > 4$. Due to this simplification, we restrict the present analysis to $D>4$ and defer the discussion of the special cases $D=3$ and $D=4$ to a dedicated section. 

 In order to stress the relevant piece of physical information following from our choices of the falloffs, we first consider  the leading-order terms  in the equations of motion $\mathcal G_\mu=0$  and analyse the outcome for the two options $b = a+1$ and $b=a$. As a result, in particular, 
a radiation solution $(b = a +1)$ is characterised by
\be
\mathcal A_u = A_u\, r^{-n/2}\,,\qquad \mathcal A_i = A_i\, r^{1-n/2}\,,
\ee
where the $r-$independent components of the potential satisfy
\be\label{radiation_rel}
A_u=\frac{2}{n}\,D^i A_i\,,
\ee
while,
on the other hand, a Coulomb-type solution $(b=a)$ is such that
\be
 {\mathcal A_u} = \tilde A_u\, r^{1-n}\,,\qquad {\mathcal A_i} = \tilde A_i\, r^{1-n}\,,
\ee
and obeys
\be\label{Coulomb_rel}
\partial_u \tilde A_u=0\,,\qquad \partial_u \tilde A_i = \frac{1}{n}\,D_i \tilde A_u\,.
\ee

Let us stress that the presence of two distinct ``branches'' of solutions, radiation and Coulombic, is apparent only for $D>4$, while in the four-dimensional case they effectively coincide. Notice also that, thanks to the condition \eqref{<0}, the nonlinear terms do not appear in these leading-order equations for $D>4$.
We are now in the position to further justify the names we gave to these two kinds of solutions: a Coulomb solution has a generically non-zero colour charge at retarded time $u$, 
\be\label{color_Coulomb}
Q^A(u)=(n-1)\int_{S_u} \mathrm{tr}\left(\tilde A_u T^A \right) d\Omega_n\,,
\ee
whereas its energy flux across $S_u$ goes to zero, due to $\mathcal F_{\, u i}  \sim r^{\, 1-n}$.
On the other hand, a radiation solution emits nonzero power at null infinity,
\be
\mathcal P(u)=-\int_{S_u} \gamma^{ij}\, \mathrm tr\left(\partial_u A_i \partial_u A_j\right)d\Omega_n\,.
\ee

At this point, two general issues are in order. To begin with, it should be stressed that the colour charge of a radiation solution diverges off-shell like $r^{n/2-1}$ as $r\to\infty$. However, employing {the relation}  \eqref{radiation_rel} and recalling that the integral of any $n$-divergence $D_i v^i$  on $S_u$ is zero by Stokes' theorem, we see that, at~least to leading order, this potentially dangerous contribution vanishes on-shell. Performing a more detailed analysis, in the next section, we will prove that these kinds of cancellations ensure the finiteness of the colour charge to all orders.

 In addition, one ought to study the behaviour of the colour flux for large $r$, namely the interplay occurring between radiation and Coulomb solutions due to the nonlinear nature of the theory. To~do~so, since the information on the colour charge is stored at order $r^{1-n}$ in the $\mathcal A_u$ component, whereas radiation contributes at order $r^{-n/2}$ in the same component, we need to consider an expansion in $1/r$ that bridges the gap between these asymptotic behaviours. Due to the appearance of a half-integer exponent, the situation changes depending on the parity of the spacetime dimension, thus justifying to differentiate the discussion into two sections. 

%%%%%%%%%%%%%%%%%%%%%%%%%%%%%%%%%%%%%%%%%%
\subsubsection{Even Space-Time Dimension}
%%%%%%%%%%%%%%%%%%%%%%%%%%%%%%%%%%%%%%%%%%

When $D > 4$ is even, we can consider an expansion of the following type:
\be\label{expansion_D_even}
\mathcal A_u = \sum_{J=1}^\infty a^{(J)} r^{1-n/2-J}\,,\qquad
\mathcal A_i = \sum_{K=0}^\infty C_i^{(K)} r^{1-n/2-K}\,,
\ee
where $a^{(J)}$ and $C_i^{(K)}$ are $r$-independent functions. On the basis of the previous discussion, we expect 
\be
a^{(1)} = A_u\,,\qquad C_i^{(0)} = A_i\, ,
\ee
to play the role of radiation terms, and 
\be
a^{(n/2)} = \tilde A_u\,, \qquad C_i^{(n/2)} = \tilde A_i\, ,
\ee
to represent the Coulomb part of the solution. The components of the field strength are then given by:
\be\begin{aligned}
	\mathcal F_{ur}&=\sum_{J=1}^\infty \left(\frac{n}{2}-1+J\right) a^{(J)} r^{-n/2-J}\,,\label{Fureven}\\[5pt]
	\mathcal F_{ir}&=\sum_{K=0}^\infty \left(\frac{n}{2}-J+K\right) C_i^{(K)} r^{-n/2-K}\,,\\[5pt]
	\mathcal F_{ui}&=\partial_u A_i\, r^{1-n/2}+ \sum_{J=1}^\infty \left(\partial_u C_i^{(J)}-D_i a^{(J)}\right) r^{1-n/2-J}+\sum_{J=1}^\infty  B_{ui}^{(J)} r^{2-n-J}\,,\\[5pt]
	\mathcal F_{ij}&=\sum_{K=0}^\infty \left(D_i C_j^{(K)}-D_j C_i^{(K)}\right) r^{1-n/2-K}+\sum_{K=0}^\infty  B_{ij}^{(K)} r^{2-n-K}\,,
\end{aligned}
\ee
where $B_{ui}^{(J)}$ and $B_{ij}^{(K)}$ contain the nonlinear terms,  
\be
B_{ui}^{(J)} := \sum_{L=1}^J\left[a^{(L)}, C_i^{(J-L)}\right] ,\qquad
B_{ij}^{(K)} := \sum_{L=0}^K\left[C_i^{(L)}, C_j^{(K-L)}\right] .
\ee	
We first substitute this expansion into the equation $\mathcal G_{r}=0$:
denoting
\be
\mathcal G_r^{(J)} := \left(\frac{n}{2}-J\right)\left(\frac{n}{2}+J-1\right)a^{(J)}-\left(\frac{n}{2}+J-2\right) D^i C_i^{(J-1)}\,,
\ee
this yields
\begin{align}
\mathcal G_r^{(J)}=0\qquad &\text{for }J=1,2,\ldots,\frac{n}{2}-1 \label{Grj0}\, ,\\
\mathcal G_r^{(J)}-\sum_{L=0}^{J-n/2}\left(\frac{n}{2}-1+L\right)\gamma^{ij}\left[C_i^{(J+n/2-L)}, C_j^{(L)}\right]=0\qquad &\text{for }J=\frac{n}{2},\frac{n}{2}+1\ldots\ \ . \label{Grjsum}
\end{align}
Then, we insert our expansion into the equation $\mathcal G_{u}=0$: 
setting
\begin{align}
\mathcal G_{u}^{\, (J)}&:=\left(\frac{n}{2}+J\right)\partial_u a^{J+1}+\left[\left(J-\frac{n}{2}\right) \left(\frac{n}{2}-1+J\right) + \Delta \right]a^{(J)}-D^i C_i^{(J)}\,,\\[10pt]
\widehat{\mathcal G}_{u}^{\, (J)}&:= D^iB_{iu}^{(J)}-\gamma^{ij}\left[C_i^{(J)},\partial_u A_j\right] \\ \nonumber
&-\sum_{L=1}^J\left\{\left(\frac{n}{2}-1+L\right)\left[a^{(J+1-L)},a^{(L)}\right]+\gamma^{ij} \left[C_i^{(J-L)},\partial_u C_j^{(L)}-D_j a^{(L)}\right] \right\} ,
\end{align} 
we obtain
%
%\begingroup\allowdisplaybreaks
\begin{align}
\frac{n}{2}\,\partial_u A_u - D^i A_i=0\, ,&\label{du1}\\[3pt]
\mathcal G_{u}^{\, (J)}=0 \, ,&\qquad \text{for }J=1,2,\ldots,\frac{n}{2}-2\label{du2}\, ,\\[5pt]
\mathcal G_{u}^{\, (n/2-1)}-\gamma^{ij}[A_i, \partial_u A_j]=0\, ,&\label{du3}\\[8pt]
\mathcal G_{u}^{\, (J)}+\widehat{\mathcal G}_{u}^{\, (J-n/2+1)}=0\, ,&\qquad \text{for }J=\frac{n}{2},\frac{n}{2}+1,\ldots,n-2\, ,\\
\mathcal G_{u}^{\, (J)}+ \widehat{\mathcal G}_{u,1}^{\, (J-n/2+1)} - \gamma^{ij}\sum_{L=1}^{J-n+2}\left[ C^{(J-n+2-L)}_i,C^{(L)}_j\right]=0\, ,&\qquad \text{for }J=n-1,n,\ldots
\end{align}
%\endgroup
%
(when $D=6$, eq.~\eqref{du2} reduces to \eqref{du1}).
It should be emphasised that the decoupling of the nonlinear terms, namely the linearity of eqs.~\eqref{Grj0}, \eqref{du1} and \eqref{du2} is a direct consequence of {the assumptions} \eqref{<0} and only holds for $D>4$. This asymptotic linearisation tells us that it is consistent to choose as boundary conditions near null infinity the falloffs \eqref{expansion_D_even}, constrained by the linear equations \eqref{Grj0}, \eqref{du1} and \eqref{du2}.
 Indeed, this set of equations will allow us to then discuss the behaviour of the charges, and its main features are the following. First, from \eqref{Grj0}, we obtain the constraints 
\be\label{constr_even}
a^{(J)}=\frac{2(n+2J-4)}{(n-2J)(n+2J-2)}\,D^i C_i^{(J)}\qquad \text{for }J=1,2,\ldots, \frac{n}{2}-1\,,
\ee
namely that $a^{(1)} ( =A_u), a^{(2)},\ldots, a^{(n/2-1)}$ are functions of the type $D_i v^i$, i.e.,\ $n$-divergences, whereas, from \eqref{Grjsum} evaluated for $J=n/2$, we note that $\tilde A_u = a^{(n/2)}$ does not bear the same form. Furthermore, \eqref{du1} and \eqref{du2} together with \eqref{constr_even} establish that  $\partial_u A_u, \partial_u a^{(2)},\ldots, \partial_ua^{(n/2-1)}$ are $n$-divergences as well. On the other hand, by \eqref{du3} and \eqref{constr_even}, 
\be\begin{aligned}\label{flux_even}
\partial_u \tilde A_u  
&=\frac{1}{n-1}\, \gamma^{ij}\left[A_i, \partial_u A_j\right] + \frac{1}{n-1} \left(D^i \tilde A_i + (n-2) a^{(n/2-1)}- \Delta a^{(n/2-1)}\right)\\
&=\frac{1}{n-1}\, \gamma^{ij}\left[A_i, \partial_u A_j\right]+(n\text{-divergence})\,.
\end{aligned}
\ee
This equation allows one to compute the evolution of the leading Coulomb term $\tilde A_u$ along the $u-$direction in terms of the leading radiation terms $A_i$, and will therefore be at the basis of our colour flux formula across $S_u$.

%%%%%%%%%%%%%%%%%%%%%%%%%%%%%%%%%%%%%%%%%%
\subsubsection{Odd Space-Time Dimension}
%%%%%%%%%%%%%%%%%%%%%%%%%%%%%%%%%%%%%%%%%%

In the case of odd dimensions $D>4$, we have to include two distinct expansions in $1/r$ in order to capture both radiation and Coulombic terms:
\be\begin{aligned}
	\mathcal A_u &= \sum_{J=1}^\infty a^{(J)} r^{1-n/2-J}
				   +\sum_{K=0}^\infty \tilde a^{(K)} r^{1-n-K}\,,\\
	\mathcal A_i &= \sum_{K=0}^\infty C_i^{(K)} r^{1-n/2-K}
				   +\sum_{K=0}^\infty \tilde C_i^{(K)} r^{1-n-K}\,,
\end{aligned}
\ee
where we identify
\be
a^{(1)} = A_u\,,\qquad C_i^{(0)} = A_i\, ,
\ee
and
\be
\tilde a^{(0)} = \tilde A_u\,,\qquad \tilde C_i^{(0)} = \tilde A_i\,.
\ee

The relevant components of the field strength are then
\be\begin{aligned}
	\mathcal F_{ur}&=\sum_{J=1}^\infty \left(\frac{n}{2}-1+J\right) a^{(J)} r^{-n/2-J} + \sum_{K=0}^\infty \left(n-1+K\right) \tilde a^{(K)}r^{-n-K}\,,\label{Furodd}\\[5pt]
	\mathcal F_{ir}&=\sum_{K=0}^\infty \left(\frac{n}{2}-1+K\right) C_i^{(K)} r^{-n/2-K} + \sum_{K=0}^\infty \left(n-1+K\right) \tilde C_i^{(K)}r^{-n-K}\,.
\end{aligned}\ee

Likewise, the equations of motion will also contain two expansions in $1/r$: one in integer powers and one in half-integer powers.
Expanding the equation $\mathcal G_r=0$, we see that
\be\label{constr_odd}
\left(\frac{n}{2}-J\right)\left(\frac{n}{2}-1+J\right) a^{(J)} - \left(\frac{n}{2}-2+J\right) D^i C_i^{(J)}=0\qquad \text{for }J=1,2,\ldots,n-1 \, ,
\ee
while the terms containing $\tilde A_u =  \tilde a^{(0)}$ cancel out identically. 

 Thus, in particular, the functions $a^{(1)} ( =A_u), a^{(2)}, \ldots, a^{(n-1)/2}$ are $n$-divergences.
Finally,~the~$r^{-n}$ order of the equation $\mathcal G_u=0$ provides us with the evolution of $\tilde A_u$ along the $u$ direction for large~$r$,~namely
\be\label{flux_odd}
(n-1)\,\partial_u \tilde A_u = \gamma^{ij}\left[A_i, \partial_u A_j\right] .
\ee

Thus, we see that the phenomenon of asymptotic linearisation of the equations of motion, emphasised in the previous section for even-dimensional spacetimes, also occurs for odd dimensions and allowed us to derive the relevant set of boundary conditions for the definition of charge and energy flux integrals.

%%%%%%%%%%%%%%%%%%%%%%%%%%%%%%%%%%%%%%%%%%
\subsubsection{Three and Four Space-Time Dimensions}\label{sec:3-4dim_1}
%%%%%%%%%%%%%%%%%%%%%%%%%%%%%%%%%%%%%%%%%%

 In $D = 4$, i.e.,\ when $n=2$, the leading radiation term and the Coulombic term coincide: indeed, finiteness of {the energy flux} \eqref{power_u} requires that, to leading order, a radiation solution behave like
\be \label{Ai34}
\mathcal A_i (u,r,x^1,x^2)= A_i(u,x^1,x^2)+\mathcal O(r^{-1})\, ,
\ee
while, using  $b = a+1$, we see that
\be \label{Au34}
\mathcal A_u (u,r,x^1,x^2) = \frac{A_u}{r} (u,x^1,x^2)+\mathcal O(r^{-2})\, .
\ee 

This also gives generically a non-vanishing colour charge on the surface $S_u$ via \eqref{color_u}. Using the leading terms in \eqref{Ai34} and \eqref{Au34}, the only relevant dynamical information arises from $\mathcal G_u=0$, which~gives
\be\label{flux_four}
\partial_u A_u = \partial_u D^i A_i + \gamma^{ij}\left[A_i, \partial_u A_j\right] .
\ee

The situation for  $D = 3$ ($n=1$) is rather different with respect to the previous cases, mainly~because of two features. First, the factor of $r^{-1}$ in \eqref{power_u} tells us that, in order to produce a~finite energy flux across $S_u$, the field components need not necessarily decay at infinity; consequently, one expects no clear distinction between radiation and Coulomb terms in the solution because no asymptotic linearisation occurs in the equations of motion.
Second, the expression \eqref{color_Coulomb}, and more specifically the factor of $r$, suggests that $\mathcal A_u$  should behave as 
$\log\frac{1}{r}$
in order to give a non-vanishing colour charge. 
These considerations motivate the following leading-order ansatz in three dimensions:
\be
\mathcal A_u (u,r,\phi) \sim  q \log \frac{1}{r}+p\,,\qquad
\mathcal A_\phi (u,r,\phi) \sim \frac{\sqrt r}{\log r}\, C\,,
\ee
where $q$, $p$ and $C$ are $r$-independent functions. Indeed, with this choice, the colour flux and the energy flux read
\be \label{QP3}
Q^A (u) = \int_{S_u} q^A d\phi\,,\qquad
\mathcal P (u)  = -\int_{S_u} \mathrm{tr}([q,C][q,C]) d\phi\, .
\ee

Using this ansatz, we find that the equation $\mathcal G_r=0$  is identically satisfied at the leading order $r^{-2}$, whereas the equation $\mathcal G_u=0$  gives
\be\label{flux_three}
\partial_u q =- \left[q,p\right] ,
\ee
at order $r^{-1}$. This equation describes the $u$-evolution of $q$ at null infinity and, hence, together with the first {formula in} \eqref{QP3}, will lead to a formula for the colour flux.
%

%%%%%%%%%%%%%%%%%%%%%%%%%%%%%%%%%%%%%%%%%%
\subsection{Asymptotic Symmetries and Charges}\label{subsec: Charges}
%%%%%%%%%%%%%%%%%%%%%%%%%%%%%%%%%%%%%%%%%%

In this section, we would like to discuss the form \eqref{color_u} of the colour charge at null infinity in the various dimensions. For related analyses, see  \cite{abbott-deser, Barnich-Brandt, Avery-Schwab, Mao_note}. 
 To begin with, let us discuss which large gauge symmetries are admissible at null infinity. The residual gauge symmetry within the radial gauge is parameterised by an $r$-independent gauge parameter, since
\be
0=\delta_\epsilon A_r = \nabla_{\!r}\, \epsilon+ \left[\mathcal A_r,\epsilon\right] 
\ee
but $\mathcal A_r=0$, hence $\nabla_{\!r}\epsilon=0$. Then, we look for those parameters $\epsilon$ that preserve the leading falloff conditions imposed on the field $\mathcal A_\mu$. In the spirit of our previous illustration, we proceed by distinguishing the case of $D>4$ from those of $D=4$ and $D=3$.

When $D>4$, where radiation gives the dominant behaviour at infinity, we find, to leading order
\beal
r^{\, - n/2} \, \delta_\epsilon A_u =  \partial_u  \epsilon + r^{-n/2}[A_u, \epsilon]\, ,
\eeal
which requires $\partial_u\epsilon=0$. Furthermore,
\be
r^{\, 1- n/2} \,\delta_\epsilon A_i = \partial_i  \epsilon+ r^{1-n/2}[A_i, \epsilon]\, ,
\ee
but, since $1-n/2<0$, this implies $\partial_i\epsilon=0$. This means that $\epsilon$ is simply a constant. Hence, in $D>4$, asymptotic symmetries coincide with the global part of the gauge group and the asymptotic charge is the ordinary colour charge computed via \eqref{color_u}.
For even space-time dimensions, using \eqref{Fureven}
\beal
Q^A(u) 
&= \lim_{r\to\infty }\int_{S_u} \mathrm{tr}(\mathcal F_{ur} T^A)\, r^{n} d\Omega_n \\
&= \lim_{r\to\infty}\sum_{J=1}^\infty r^{n/2-J} \left(\frac{n}{2}-1+J\right) \int_{S_u} \mathrm{tr}(a^{(J)}T^A)\, d\Omega_n\, ,
\eeal
where for  $J<n/2$ all terms are integrals of $n$-divergences thanks to {the relation} \eqref{constr_even}, while the terms with $J>n/2$ go to zero as $r\to\infty$, thus
\be\label{ColoreD>4}
Q^A(u)=(n-1)\int_{S_u} \mathrm{tr}(\tilde A_u T^A)\, d\Omega_n\, .
\ee

For odd space-time dimensions,
\beal
Q^A(u) 
&= 
\lim_{r\to\infty}\sum_{J=1}^\infty r^{n/2-J} \left(\frac{n}{2}-1+J\right) \int_{S_u} \mathrm{tr}(a^{(J)}T^A)\, d\Omega_n\\
&+
\lim_{r\to\infty}\sum_{K=0}^\infty r^{-K} \left(n-1+K\right) \int_{S_u} \mathrm{tr}(\tilde a^{(K)}T^A)\, d\Omega_n\,,
\eeal
and, by {the relation} \eqref{constr_odd}, the only nonzero contribution comes from the $K=0$ term of the second series, giving again {the result} \eqref{ColoreD>4}. For $D>4$, we thus obtained that the colour charge is indeed expressed as an~integral of the leading Coulombic component on $S_u$. Furthermore, on account of \eqref{flux_even} and \eqref{flux_odd}, the colour flux is written as
\be
\frac{d}{du} Q^A(u) = \int_{S_u} \gamma^{ij} \left[A_i, \partial_u A_j\right]^A d\Omega_n\,.
\ee
This is indeed consistent with the interpretation of $A_i$ as the leading radiation term: this formula describes how Yang--Mills radiation across null infinity induces a change in the total colour of the space-time at successive retarded times $u$.

In $D=4$, the gauge parameter must satisfy:
\beal
r^{\, -1} \, \delta_\epsilon A_u & = \partial_u  \epsilon + r^{-1}[A_u, \epsilon] \, ,\\[5pt]
\delta_\epsilon A_i & = \partial_i  \epsilon + [A_i, \epsilon] \, .
\eeal
The first equation again enforces $\partial_u \epsilon=0$, whereas the second allows for an $\epsilon(x^1, x^2)$ with arbitrary dependence on the angles on the celestial sphere. The corresponding asymptotic charge is therefore
\be
Q_\epsilon(u) = \lim_{r\to\infty}\int_{S_u} \mathrm{tr}(\mathcal F_{ur}\epsilon)\, r^2 d\Omega_2 = \int_{S_u} \mathrm{tr}(A_u \epsilon)\, d\Omega_2\,.
\ee
Taking into account \eqref{flux_four},
\be
\frac{d}{du} Q_\epsilon(u) = \int_{S_u} \mathrm{tr}\left[\big(\partial_u D^i A_i+\gamma^{ij}[A_i, \partial_u A_j]\big) \epsilon\right] d\Omega_2\,.
\ee

 To complete the picture, let us now turn to the situation in $D=3$. There, neither $\mathcal A_u$ nor $\mathcal A_\phi$ fall off at infinity, and hence any $\epsilon(u,\phi)$ generates an allowed gauge transformation (the same~result, in~a~slightly different setting, was already obtained in \cite{Maxwell d=3 Barnich}). Thus, using the notation of the previous~section,
\beal
Q_\epsilon(u) &= \int_{S_u} \mathrm{tr}(q \epsilon) d\phi\,,\\
\frac{d}{du}Q_\epsilon(u) &= \int_{S_u} \mathrm{tr}(q \partial_u\epsilon) d\phi - \int_{S_u} \mathrm{tr}([q, p] \epsilon )d\phi\,.
\eeal

Let us observe that these charges indeed form a representation of the underlying algebra: for~$D\ge4$, since $\delta_\epsilon A_u= [A_u,\epsilon]$,
\be\label{commutation_rel}
[Q_{\epsilon_1},Q_{\epsilon_2}]=\delta_{\epsilon_1} Q_{\epsilon_2}= \int_{S_u} \mathrm{tr}([A_u, \epsilon_1]\epsilon_2) d\Omega_n =  \int_{S_u} \mathrm{tr}(A_u [\epsilon_1,\epsilon_2]) d\Omega_n= Q_{[\epsilon_1,\epsilon_2]}\,;
\ee
the same result holds for $D=3$, noting that $\delta_\epsilon q= [q,\epsilon]$ and $\delta_\epsilon p = \partial_u \epsilon$, but $p$ does not enter the charge formula.  While {the identity} \eqref{commutation_rel} holds in any dimension, it should be stressed that, when $D>4$, the corresponding charge algebra coincides with $\mathfrak g$, whereas in $D=4$ and $D=3$, it is in fact an infinite-dimensional Kac--Moody algebra, owing to the arbitrary gauge parameters $\epsilon(x^1,x^2)$ and $\epsilon(u,\phi)$. In particular, we~note the absence of a central charge, which could however emerge by performing the analysis for the linearised theory around a nontrivial background, as pointed out in \cite{Avery-Schwab}.

 Let us conclude this section by presenting some general observations that, although of basic~nature, we found useful in order to frame the correctness of our procedure. For Yang--Mills theory, the following~quantity
\be \label{charge_general}
Q_\epsilon= \int_{\partial \Sigma} dx_{\mu\nu}\, \mathrm{tr}\left(F^{\mu\nu} \epsilon\right) ,
\ee
where $\Sigma$ is a generic Cauchy surface, provides both the conserved charge, as obtained by the Noether algorithm, and the Hamiltonian generator of the gauge symmetry parameterised by $\epsilon$ on the space tangent to the surface of solutions, as calculated via covariant phase space methods. 
Indeed, a generic variation of the Yang--Mills Lagrangian, after integrating by parts, reads
\be\label{delta L}
\delta \mathcal L=-\,\mathrm{tr}\left(\mathcal G^{\mu} \delta A_\mu\right) + \partial_\mu \mathrm{tr}\left(F^{\mu\nu}\delta A_\nu\right) =: -\mathrm{tr}\left(\mathcal G^{\mu} \delta A_\mu\right) +\partial_\mu \theta^\mu(\delta A) \,,
\ee
where we defined the symplectic potential $\theta^\mu(\delta A)=\mathrm{tr}\left(F^{\mu\nu} \delta A_\nu\right)$, while $\mathcal G^\mu$ denotes the Euler--Lagrange derivatives of $\mathcal L$, given in \eqref{EEOM}. The presymplectic form is then given by 
\be
\omega^{\mu}(\delta_1 A, \delta_2 A)=\delta_{[1}\theta^\mu(\delta_{2]}A)\, ,
\ee
with square brackets denoting antisymmetrisation, and correspondingly the formal variation of the Hamiltonian generator of the gauge symmetry  
$H_\epsilon$ is
\be
\slashed{\delta} H_\epsilon = \int_\Sigma dx_\mu \omega^{\mu}(\delta A, \delta_\epsilon A)=\delta \int_{\partial\Sigma} dx_{\mu\nu} \mathrm{tr}(F^{\mu\nu}\epsilon)- \int_{\Sigma} dx_\mu \mathrm{tr}(\delta\mathcal G^\mu \epsilon)\,.
\ee
Noting that the last term is proportional to the linearised equations of motion, i.e.,\ that it vanishes on the space tangent to the surface of solutions, we can write
$$
\slashed{\delta} H_\epsilon \approx \delta Q_\epsilon\,,
$$
which explicitly shows that $\slashed{\delta} H_\epsilon$ is integrable and that we may choose to set $H_\epsilon = Q_\epsilon$ by requiring a~flat connection to have zero colour charge. 
Furthermore, the Noether charge is simply
\be
\int_{\Sigma}dx_\mu\, \theta^{\mu}(\delta_\epsilon A) = Q_\epsilon - \int_\Sigma dx_\mu\, \mathrm{tr}(\mathcal G^\mu \epsilon) \approx Q_\epsilon\,,
\ee
so that the two approaches agree in this case. The definition of $Q_\epsilon$ is in principle subject to ambiguities stemming from $\theta^\mu \mapsto \theta^\mu + \partial_\nu \lambda^{\mu\nu}$, where $\lambda^{\mu\nu}=-\lambda^{\nu\mu}$, which does not alter {the variation} \eqref{delta L}. In the spirit of \cite{Wald-Zoupas}, we may choose to set to zero the corresponding additional terms, precisely because this choice defines an integrable Hamiltonian, as shown above. Further motivation for the absence of these terms is provided by the agreement with the general analysis of \cite{Barnich-Brandt} and by the fact that they play no role in the generation of Ward identities for residual gauge freedom \cite{Avery-Schwab}.

 In order to finally make contact with \eqref{color_u}, we may then apply \eqref{charge_general} choosing as a Cauchy surface 
$$
\Sigma = \Sigma_u \cup \mathscr{I}^+_{< u} \, ,
$$
where $\Sigma_u$ is any space-like hypersurface such that $\partial \Sigma_u = S_u$ while $\mathscr{I}^+_{< u}$ is the portion of null infinity up to the retarded time $u$. Then, using {the general expression for the charge} \eqref{charge_general} and Stokes'  theorem, we see that $Q_\e$ can be expressed as a sum of the total charge at the retarded time $u$, as in \eqref{color_u}, and the charge flown across $\mathscr{I}^+_{< u}$ due to radiation.

\vspace{6pt}
%%%%%%%%%%%%%%%%%%%%%%%%%%%%%%%%%%%%%%%%%%%%%%%%%%%%%%%%%%%%%%%%%%%%%
\section{Linearised Gravity}\label{sec:spin2}
%%%%%%%%%%%%%%%%%%%%%%%%%%%%%%%%%%%%%%%%%%%%%%%%%%%%%%%%%%%%%%%%%%%%%

Boundary conditions giving finite energy and angular momentum at null infinity have been first proposed for spacetimes of any even dimensions in \cite{gravity_evenD_1} (see also \cite{positivity,gravity_evenD_2, Pate_Memory}). The proposal has been extended to encompass also odd space-time dimensions in \cite{gravity_anyD,angular-momentum}. We refer to these works for a detailed analysis of asymptotic charges and fluxes at null infinity in nonlinear Einstein gravity.  Here, we revisit instead the problem within the linearised theory. In particular, we point out that the boundary conditions discussed in previous works can be inferred by demanding finiteness of the linearised asymptotic charges. In analogy with the Yang--Mills example, the fluxes of energy and angular-momentum at null infinity are instead affected by interactions, so that they will be excluded from our analysis. Besides its intrinsic interest, the ensuing discussion is also instrumental for us in order to better frame the results that we will present for higher-spin fields in Sections~\ref{sec:spin3} and  \ref{sec:spin-s}.

%%%%%%%%%%%%%%%%%%%%%%%%%%%%%%%%%%%%%%%%%%
\subsection{Boundary Conditions}
%%%%%%%%%%%%%%%%%%%%%%%%%%%%%%%%%%%%%%%%%%

We parameterise the Minkowski background with the retarded Bondi coordinates \eqref{bondi-coord}, and~we~analyse the linearised metric fluctuations in the ``Bondi gauge''
\be \label{bondi2}
h_{r\m} = 0 \, , \quad \g^{ij} h_{ij} = 0 \quad \Rightarrow \quad g^{\m\n} h_{\m\n} = 0 \, .
\ee
Differently from {the spin-one radial gauge} \eqref{bondi1}, these conditions cannot be reached by means of an off-shell gauge fixing, but~the number of constraints is the same as in the transverse-traceless gauge. We therefore assume that they can be imposed on shell.\footnote{With hindsight, our choice is legitimated, e.g., by the agreement between the charges and asymptotic symmetries derived in this framework and those obtained by assuming only suitable falloff conditions on the components of the metric that cannot be set to zero with an off-shell gauge fixing (compare e.g.\ {the conditions} \eqref{bondi2} with eq.~(8) of \cite{gravity_evenD_1}). }

When \eqref{bondi2} holds, the linearised vacuum Einstein equations reduce to
\be \label{Ricci}
R_{\m\n} = \Box h_{\m\n} - \nabla_{\!(\m} \nabla\cdot h_{\n)} = 0 \, .
\ee

In the following, we will solve these equations assuming that the metric fluctuations admit an~expansion in powers of $1/r$ around null infinity. The main idea, suggested by the Yang--Mills example, is that asymptotically the interactions deform the linearised solutions only starting from a~subleading order in their expansion in powers of $1/r$. 
The conditions \eqref{bondi2} imply $R_{rr} = 0$ identically, while the other equations of motion read\footnote{From now on, we shall often denote a derivative with respect to $u$ with a dot, i.e.,\ $\pr_u f = \dot{f}$.}
\begin{align}
R_{ru} & = \frac{1}{r^2} \left\{ \left( r^2 \pr_r^2 + n\, r\pr_r \right) h_{uu} - \pr_r \Ddot h_u \right\} = 0 \, , \label{Rru} \\[10pt]
R_{ri} & = \frac{1}{r^2} \left( r^2 \pr_r^2 + (n-2)\, r\pr_r - 2(n-1) \right) h_{ui} - \frac{1}{r^3} \left( r\pr_r - 2 \right) \Ddot h_i = 0 \, , \label{Rri} \\[10pt]
R_{ij} & = -\, \frac{1}{r} \left( 2\,r\pr_r + n - 4 \right)\dot{h}_{ij} + \frac{1}{r} \left\{ \left( r\pr_r + n - 2 \right) D_{(i} h_{j)u} + 2\, \g_{ij} \Ddot h_u \right\} \nn \\
& + \frac{1}{r^2} \left\{ \left( \Delta + r^2 \pr_r^2 + (n-4)\,r\pr_r - 2(n-2) \right) h_{ij} - D_{(i} \Ddot h_{j)}  \right\} - 2\, \g_{ij} \left( r\pr_r + n - 1 \right) h_{uu} = 0 \, . 
\label{Rij}
\end{align}
When the previous equations are satisfied, the following ones are satisfied as well at almost all orders in an expansion in powers of $1/r$ (see Section~\ref{sec:spin-s} for more details): 
\begin{align}
R_{uu} & = \frac{n}{r}\, \dot{h}_{uu} - \frac{2}{r^2}\, \Ddot \dot{h}_u + \frac{1}{r^2} \left( \Delta + r^2\pr_r^2 + n\,r\pr_r \right) h_{uu} = 0 \, , \label{Ruu} \\[10pt]
R_{ui} & = -\, \frac{1}{r} \left( r\pr_r -2 \right) \dot{h}_{ui} - \frac{1}{r^2}\, \Ddot \dot{h}_i + \frac{1}{r} \left( r\pr_r + n - 2 \right) \pr_i h_{uu} \nn \\
& + \frac{1}{r^2} \left\{ \left( \Delta + r^2 \pr_r^2 + (n-2)\, r\pr_r - n + 1 \right) h_{ui} - D_i \Ddot h_u \right\} = 0 \, .\label{Rui}
\end{align}
As for the Yang--Mills case, the only exception is given by the leading order of a stationary solution.

By substituting a power-law ansatz,
\be
h_{uu} = r^a B(u,x^k) + \cO(r^{a-1}) \, , \quad
h_{ui} = r^b U_i(u,x^k) + \cO(r^{b-1}) \, , \quad
h_{ij} = r^c C_{ij}(u,x^k) + \cO(r^{c-1}) \, ,
\ee
eqs.~\eqref{Rru}--\eqref{Rij} turn into
%
%\begingroup\allowdisplaybreaks
\begin{align}
R_{ru} & = r^{a-2} a(a+n-1) B - r^{b-3} b\, \Ddot U + \cdots = 0  \, , \label{Rru_lead} \\[10pt]
R_{ri} & =  r^{b-2} (b-2)(b+n-1)\, U_i - r^{c-3} (c-2) \Ddot C_i + \cdots = 0 \, , \label{Rri_lead} \\[10pt]
\begin{split}
R_{ij} & = -\, r^{c-1} (2c+n-4)\, \dot{C}_{ij} + r^{b-1} \left\{ (b+n-2) D_{(i}U_{j)} + 2\,\g_{ij} \Ddot U \right\} \\*
& - 2\,r^a (a+n-1) \g_{ij} B + \cdots = 0 \, , \label{Rij_lead}
\end{split}
\end{align}
%\endgroup
%
where the dots stand for subleading terms. Imposing $b= a + 1$ and $c = b + 1$ allows one to mutually cancel  the addenda in \eqref{Rru_lead} and \eqref{Rri_lead}, while \eqref{Rij_lead} is solved to leading order provided that the coefficient of $r^{c-1}$ vanishes. This is the analogue of the choice that gives the radiation solution in the Yang--Mills case: it does not impose any constraint on $C_{ij}$ while, for $D > 3$, it fixes the leading exponents as follows:
\be \label{leading-falloff_2}
a = -\, \frac{n}{2} \, , \qquad
b = -\, \frac{n}{2} + 1 \, , \qquad
c = -\, \frac{n}{2} + 2 \, .
\ee

Besides this formal analogy, one can verify that a solution of this type carries a finite amount of energy per unit of retarded time through null infinity, \footnote{The massless Fierz--Pauli Lagrangian 
$\mathcal L= \frac{1}{2}  h^{\mu\nu}(\Box h_{\mu\nu} - \nabla_{\!(\mu}\nabla\cdot h_{\nu)} + \nabla_{\!\mu} \nabla_{\!\nu} h\indices{^\alpha_\alpha}-\eta_{\mu\nu} (\Box h\indices{^\alpha_\alpha}-\nabla\cdot\nabla\cdot h))$
gives rise, in Bondi gauge, to the canonical stress-energy tensor 
$$
T_{\alpha\beta}= \nabla_{\!\alpha} h_{\mu\nu} \nabla_{\!\beta} h^{\mu\nu} -2 \nabla\cdot h^\mu \nabla_{\!\alpha} h_{\beta\mu}+\eta_{\alpha\beta}\mathcal L \, .
$$
%, 
While in our linearised setup one cannot capture the flux of energy associated with the self-interactions of the gravitational field, it still makes sense to evaluate the flux pertaining to an {\it eternal} radiating source in the interior, which is constant over $u$. Indeed, this is a quantity that is well defined also in the linearised theory and is given by \eqref{energy-flux2}.}
\be \label{energy-flux2}
\cP(u) = \lim_{r\to \infty} \int_{S_u} \left(T_{uu}-T_{ur}\right) r^n d\Omega_n = \int_{S_u} \gamma^{i_1j_1}\gamma^{i_2 j_2}\,\dot{C}_{i_1i_2} \dot{C}_{j_1 j_2}d\Omega_n \, ,
\ee
and can therefore be interpreted as a gravitational wave propagating on the Minkowski background, thus providing a convincing justification for the falloffs  \eqref{leading-falloff_2}. In Section~\ref{sec:charges2}, we shall also show that a solution of the Einstein equations with these leading falloffs is endowed with finite energy and angular momentum charges at null infinity. 

The charges actually depend on the subleading (for $D > 3$) terms in the expansion in powers of $r$ with exponents
\be
a = b = c = 1 - n \, .
\ee
When $n$ is even, these contributions are actually ``integration constants'' (they anyway admit a~dependence on $x^i$) in the radiation solution with leading falloffs \eqref{leading-falloff_2}, while when $n$ is odd they appear as the leading order of a companion solution with its own expansion in powers of $1/r$. For~this reason, we treat separately the two cases, while discussing the peculiarities of the $n=1$ and $n=2$ instances in Section~\ref{sec:3-4dim_2}.

%%%%%%%%%%%%%%%%%%%%%%%%%%%%%%%%%%%%%%%%%%
\subsubsection{Even Space-Time Dimension}\label{sec:even_2}
%%%%%%%%%%%%%%%%%%%%%%%%%%%%%%%%%%%%%%%%%%

When $D = n - 2$ is even, we consider the following ansatz for the linearised fluctuations in Bondi gauge \eqref{bondi2}:
\be \label{ansatz2}
h_{uu} = \sum_{k\,=\,0}^\infty r^{-\frac{n}{2}-k} B^{(k)}(u,x^m) \, , \quad
h_{ui} = \sum_{k\,=\,0}^\infty r^{-\frac{n}{2}-k+1} U^{(k)}_i(u,x^m) \, , \quad
h_{ij} = \sum_{k\,=\,0}^\infty r^{-\frac{n}{2}-k+2} C^{(k)}_{ij}(u,x^m) \, ,
\ee
with $\g^{ij} C^{(k)}_{ij} = 0$. As discussed above, the leading falloffs have been chosen such that the linearised solution carries a finite amount of energy per unit time at null infinity. From eq.~\eqref{Rru}, one can then compute the coefficients of $h_{uu}$, 
\be
B^{(k)} = 
\left\{
\begin{array}{ll}
\frac{2(n+2k-2)}{(n+2k)(n-2k-2)}\, \Ddot U^{(k)} \qquad & \textrm{for}\ k \neq \frac{n-2}{2} \label{solB2} \\[5pt]
2\,m_B & \textrm{for}\ k = \frac{n-2}{2}
\end{array}
\right. ,
\ee
while eq.~\eqref{Rri} fixes $h_{ui}$ as
\be
U^{(k)}_i = \left\{
\begin{array}{ll}
\frac{2(n+2k)}{(n+2k+2)(n-2k)}\, \Ddot C^{(k)}_i \qquad & \textrm{for}\ k \neq \frac{n}{2} \label{solU2} \\[5pt]
N_i & \textrm{for}\ k = \frac{n}{2}
\end{array}
\right. .
\ee

The quantities $m_B$ and $N_i$ do not contribute to eqs.~\eqref{Rru} and \eqref{Rri} thanks to the cancellation of the coefficients in front of the corresponding $B^{(k)}$ and $U_i^{(k)}$.  Their dependence on the retarded time $u$ is however fixed by eqs.~\eqref{Ruu} and \eqref{Rui} that, when $n>2$, read
\begin{subequations} \label{eq-u_spin2}
\begin{align}
\dot{m}_B & = \frac{3-n}{4n(n-1)} \left( \Delta - n + 2 \right) \Ddot\Ddot C^{(\frac{n-4}{2})} \, , \label{eq-uB2} \\[10pt]
\dot{N}_i & = \frac{1}{n+1} \left\{ 2\,\pr_i m_B - \frac{n-1}{n} \left( \Delta - 1 \right) \Ddot C_i^{(\frac{n-2}{2})} \right\} , \label{eq-uU2}
\end{align}
\end{subequations}
where $\Delta  = D_i D^i$. These equations are solved by
\begin{subequations}\label{int-const2}
\begin{gather}
\hspace{-24pt}m_B(u,x^j)  = \cM(x^j) - \frac{n-3}{4(n-1)n} \int_{-\infty}^u \!\!du' \left( \Delta - n + 2 \right) \Ddot\Ddot C^{(\frac{n-4}{2})}(u',x^j) \, , \label{uB2} \\
\begin{aligned}
N_i(u,x^j)  = \cN_i(x^j) + \frac{2\,u}{n+1}\, \pr_i \cM (x^j)  - \frac{n-1}{n(n+1)}  \int_{-\infty}^u \!\!du' \left( \Delta - 1 \right) \Ddot C_i{}^{\!(\frac{n-2}{2})}(u',x^j)  \\
 - \frac{n-3}{2(n-1)n(n+1)} \int_{-\infty}^u \!\!du'\! \int_{-\infty}^{u'} \!\!du''\, D_i \left( \Delta - n + 2 \right) \Ddot\Ddot C^{(\frac{n-4}{2})}(u'',x^j) \, . \label{uU2}
\end{aligned}
\end{gather}
\end{subequations}

Note that the expressions for $m_B$ and $N_i$ (which are the linearised counterparts of the Bondi mass and angular momentum aspects) contain two types of contributions: one depends on  the ``integration constants'' $\cM$ and $\cN_i$, which enter in combinations with a fixed dependence on $u$, while the other depends on the integrals over $u$ of certain combinations of the tensors $C_{ij}^{(k)}$. We anticipate that in Section~\ref{sec:charges2} we shall show that, for $n \neq 2$, the integration constants $\cM$ and $\cN_i$ completely specify the asymptotic linearised charges, while the integral terms, which are not even present when the dimension of spacetime is odd, do not contribute to them.

The tensors in the expansion of $h_{ij}$ are instead fixed recursively in terms of $C_{ij}{}^{(0)}$---whose $u$-derivative is the linearised analogue of the Bondi news---up to an arbitrary function of $x^i$ for each term of the expansion. Eq.~\eqref{Rij} indeed implies
\be \label{solC2}
\begin{split}
\dot{C}^{(k+1)}_{ij} & = -\, \frac{1}{2(k+1)}\, \bigg\{ \left[\, \Delta - \frac{n(n-2)}{4} + k(k+1) - 2 \,\right] C^{(k)}_{ij} \\
& - \frac{4}{(n+2k+2)(n-2k)} \left[\, n\, D_{(i}\Ddot C_{j)}{}^{\!\!(k)} - 2\, \g_{ij} \Ddot\Ddot C^{(k)} \,\right] \bigg\} \quad \forall\ k  \neq \frac{n}{2} \, .
\end{split}
\ee
The value of $k$ excluded from this expression shows that from $k = \frac{n+2}{2}$ onwards the tensors also depend on~$N_i$:
\be \label{solC2-special}
\dot{C}^{(\frac{n+2}{2})}_{ij} = \frac{1}{n+2} \left\{ D_{(i\,} N_{j)} - \frac{2}{n}\, \g_{ij} \Ddot N 
- \left( \Delta + n - 2 \right) C^{(\frac{n}{2})}_{ij} + D^{\phantom{(\frac{n}{2})}}_{(i} \!\!\! \Ddot C^{(\frac{n}{2})}_{j)} \right\} .
\ee
These terms in the expansion \eqref{ansatz2}, anyway, do not contribute to the linearised charges and, generically, they receive nonlinear corrections in Einstein gravity \cite{gravity_anyD,angular-momentum}.

For the values of $k$ that do not impose any constraints on $B^{(k)}$ and $U_i{}^{\!(k)}$, eqs.~\eqref{Rru} and \eqref{Rri} imply instead
\be \label{constrC2}
(n-2)(n-1)\,\Ddot \Ddot C_{\phantom{i}}^{(\frac{n-2}{2})} = 0 \, , \qquad
\Ddot C_i^{(\frac{n}{2})} = 0 \, .
\ee
These conditions do not constrain $C^{(0)}$ because they are compatible with the divergences of \eqref{solC2}. For instance, it implies
\be
\Ddot \Ddot \dot{C}^{(k+1)} = - \frac{(n+2k-2)(n-2k-4)}{2(k+1)(n+2k+2)(n-2k)} \left[ \Delta - \frac{(n+2k)(n-2k-2)}{4} \right] \Ddot\Ddot C^{(k)} 
\ee
and, for $n>2$, the r.h.s.\ vanishes for $k+1=\frac{n-2}{2}$.
Similarly, eqs.~\eqref{Ruu} and \eqref{Rui} reduce to divergences of \eqref{solC2} for all values of $k$ aside from those that fix the $u-$dependence \eqref{int-const2} of the Bondi mass and angular momentum aspects.

%%%%%%%%%%%%%%%%%%%%%%%%%%%%%%%%%%%%%%%%%%
\subsubsection{Odd Space-Time Dimension}\label{sec:odd_2}
%%%%%%%%%%%%%%%%%%%%%%%%%%%%%%%%%%%%%%%%%%

When $n$ is odd and greater than one, in order to obtain non-zero asymptotic charges at null infinity, one has to complement the ansatz \eqref{ansatz2}, that in this case contains half-integer powers of $r$, with~a companion expansion including integer powers of the radial coordinate.  We therefore consider the ansatz
\begin{subequations} \label{ansatz2-odd}
\begin{align} 
h_{uu} & = \sum_{k\,=\,0}^\infty r^{-\frac{n}{2}-k} B^{(k)}(u,x^m) + \sum_{k\,=\,0}^\infty r^{1-n-k} \tilde{B}^{(k)}(u,x^m) \, , \\
h_{ui} & = \sum_{k\,=\,0}^\infty r^{-\frac{n}{2}-k+1} U^{(k)}_i(u,x^m) + \sum_{k\,=\,0}^\infty r^{1-n-k} \tilde{U}_i^{(k)}(u,x^m) \, , \\
h_{ij} & = \sum_{k\,=\,0}^\infty r^{-\frac{n}{2}-k+2} C^{(k)}_{ij}(u,x^m) + \sum_{k\,=\,0}^\infty r^{1-n-k} \tilde{C}_{ij}^{(k)}(u,x^m)\, ,
\end{align}
\end{subequations}
with $\g^{ij} C_{ij}^{(k)} = \g^{ij} \tilde{C}_{ij}^{(k)} = 0$. Since $n$ is odd, the factors entering the expansion of \eqref{Rru} and \eqref{Rri} in powers of $\sqrt{r}$ are always different from zero. As a result, for any $k$ one has again
\be \label{solB2-odd}
B^{(k)} = \frac{2(n+2k-2)}{(n+2k)(n-2k-2)}\, \Ddot U^{(k)} \, , \qquad
U_i^{(k)} = \frac{2(n+2k)}{(n+2k+2)(n-2k)}\, \Ddot C^{(k)}_i \, ,
\ee
while the tensors $C^{(k)}$ satisfy \eqref{solC2}. These conditions imply that the equations \eqref{Ruu} and \eqref{Rui} are identically satisfied. In Section~\ref{sec:charges2}, we shall see that {the relations} \eqref{solB2-odd} guarantee that the radiation solution does not contribute to the asymptotic charges.
Eqs.~\eqref{Ruu} and \eqref{Rui} fix instead the $u$-evolution of the leading terms in the Coulomb-type solution as
\begin{subequations} \label{int-const2-odd}
\begin{align}
m_B(u,x^j) & := \frac{\tilde{B}^{(0)}}{2} = \cM(x^j) \, , \label{uB2-odd} \\[10pt]
N_i(u,x^j) & := \tilde{U}_i{}^{\!(0)} = \cN_i(x^j) + \frac{2\,u}{n+1}\, \pr_i \cM (x^j) \, . \label{uU2-odd}
\end{align}
\end{subequations}

The subleading terms in the expansion in powers of $r$ are fixed by the analogues of the relations~\eqref{solC2} and \eqref{solB2-odd}. For instance,
\be
\tilde{B}^{(l+1)} = - \frac{n+l-1}{(l+1)(n+l)}\, \Ddot \tilde{U}^{(l)} \, , \qquad
\tilde{U}_i^{(l)} = - \frac{n+l+1}{(l+1)(n+l+2)}\, \Ddot \tilde{C}^{(l)}_i \, ,
\ee
while the $\tilde{C}^{(l)}$ are fixed recursively by an equation with the same form as \eqref{solC2} with shifted coefficients $k \to k - n/2 - 1$ (see also the general analysis in Section~\ref{sec:spin-s} for more details).  
At any rate, these~relations will be irrelevant for the computation of the charges in Section~\ref{sec:charges2} and, in general, will receive nonlinear corrections in Einstein gravity as shown by the comparison of our analysis with \cite{gravity_anyD,angular-momentum}. 

%%%%%%%%%%%%%%%%%%%%%%%%%%%%%%%%%%%%%%%%%%
\subsubsection{Three and Four Space-Time Dimensions}\label{sec:3-4dim_2}
%%%%%%%%%%%%%%%%%%%%%%%%%%%%%%%%%%%%%%%%%%

In three and four space-time dimensions, i.e.,\ when $n = 1$ or $n = 2$, the previous analysis has to be amended for some details, which however introduce significant physical consequences. We begin by considering the peculiarities that emerge in four dimensions: in this case, inserting our ansatz \eqref{ansatz2} in the equations of motion \eqref{Rru} and \eqref{Rri} leads to
\be \label{falloffs4D}
h_{uu} = \frac{2}{r}\, m_B + \cO(r^{-2}) \, , \qquad
h_{ui} = \frac{1}{2}\, \Ddot C_i + \frac{1}{r}\, N_i + \cO(r^{-2}) \, , \qquad
h_{ij} = r\, C_{ij} + \cO(1) \, ,
\ee
where, for brevity, we defined $C_{ij} := C_{ij}{}^{(0)}$, while we used the same notation as in the previous subsections for the leading terms of the Coulomb-like solution.
Note that in the component $h_{uu}$ the leading order of the radiation and Coulomb-like solutions coincide. Moreover, eq.~\eqref{Rru} does not impose any constraint on the double divergence of $C_{ij}$ (the factor in front of it vanishes when $n=2$ as we recalled in \eqref{constrC2}), while $\Ddot C_{i}{}^{(1)} = 0$ as for generic $n$. Since $C_{ij}$ has now a non-vanishing double divergence, the equations fixing the dependence on $u$ of $m_B$ and $N_i$ have to be modified as follows (cf.~\eqref{eq-u_spin2}):
\begin{subequations}
\begin{align}
\dot{m}_B & = \frac{1}{4}\, \pr_u \Ddot\Ddot C \, , \label{eq1} \\[10pt]
\dot{N}_i & = \frac{2}{3}\, \pr_i m_B - \frac{1}{6} \left\{ \left( \Delta - 1 \right) \Ddot C_i - D_i \Ddot \Ddot C\, \right\} .
\end{align}
\end{subequations}
Consequently, the leading terms of the Coulomb-like branch depend on the radiation solution and on the usual set of integration constants as 
\begin{subequations} \label{int4D}
\begin{align}
m_B(u,x^j) & = \cM(x^j) + \frac{1}{4}\, \Ddot\Ddot C(u,x^j) \, , \\[10pt]
N_i(u,x^j) & = \cN_i(x^j) + \frac{2\,u}{3}\, \pr_i \cM (x^j) - \frac{1}{6}  \int_{-\infty}^u \!\!du' \left[ \left( \Delta - 1 \right) \Ddot C_i - 2\, D_i \Ddot \Ddot C \,\right]\!(u',x^j) \, .
\end{align}
\end{subequations}
The dependence on $C_{ij}$ in $m_B$ and $N_i$ will be crucial in Section~\ref{sec:charges2}, where we shall compute the asymptotic charges.

When $n=1$, in the component $h_{uu}$ the radiation branch becomes subleading with respect to the Coulomb-type one. Given that in three space-time dimensions fields of spin two do not propagate any local degrees of freedom, it is therefore natural to ignore the radiation branch altogether and work with boundary conditions that only encompass Coulomb-type solutions of the equations of~motion:
\be
h_{uu} = 2\cM(\phi) + \cO(r^{-1}) \, , \qquad
h_{u\phi} = \cN(\phi) + u\, \pr_\phi \cM(\phi) + \cO(r^{-1}) \, , \qquad
h_{\phi\phi} = 0 \, ,
\ee
where $\phi$ denotes the angular coordinate on the circle at null infinity while we already displayed the constraints on the leading terms imposed by the equations of motion. Notice that we set to zero the component $h_{\phi\phi}$, consistently with our choice of boundary conditions in any $D$ according to which  the tensor $h_{\, i j }$ is traceless (and thus identically zero if $n = 1$). Alternatively, one can consider $h_{\phi\phi} = r C(\phi) + \cO (1)$  \cite{bms3}. Our choice is not restrictive, however, as it still allows for an enhancement of the asymptotic symmetry algebra from Poincar\'e to $\mathfrak{bms}_3$.

%%%%%%%%%%%%%%%%%%%%%%%%%%%%%%%%%%%%%%%%%%
\subsection{Asymptotic Symmetries and Charges}\label{sec:charges2}
%%%%%%%%%%%%%%%%%%%%%%%%%%%%%%%%%%%%%%%%%%

We can now identify the gauge transformations preserving the form of the linearised solutions, which are the asymptotic symmetries of the system. These determine the asymptotic charges, which play the dual role of being conserved quantities labelled by the parameters of asymptotic symmetries and of generating the latter via the Poisson bracket derived from the action.
In the Bondi gauge \eqref{bondi2}, the~asymptotic symmetries must satisfy
\be \label{fix-var_2}
\d h_{r\m} = 0 \, , \quad
\d h_{uu} = \cO(r^{-\frac{n}{2}}) \, , \quad
\d h_{ui} = \cO(r^{-\frac{n}{2}+1}) \, , \quad
\d h_{ij} = \cO(r^{-\frac{n}{2}+2}) \, ,
\ee 
where the variations are given by linearised diffeomorphisms $\d h_{\m\n} = \nabla_{\!(\m} \x_{\n)}$ that leave \eqref{Ricci} invariant.
These conditions are appropriate for spacetimes of both even and odd dimensions, since the differences highlighted in Sections~\ref{sec:even_2} and \ref{sec:odd_2} only affect subleading terms in the expansion of the solutions.\footnote{The exception is given by $n = 1$. Accordingly with the discussion in Section~\ref{sec:3-4dim_2}, in three dimensions, the conditions \eqref{fix-var_2} are substituted by
\[
\d h_{uu} = \cO(1) \, , \quad
\d h_{u\phi} = \cO(1) \, , \quad
\d h_{\phi\phi} = 0 \, , \quad
\d h_{r\m} = 0 \, .
\]
}
Eqs.~\eqref{fix-var_2} are solved by 
\begin{subequations} \label{xi2}
\begin{align}
\x_r & = - \left( T + \frac{u}{n}\,\Ddot v \right) , \label{x2_r} \\
\x_u & = \frac{r}{n}\, \Ddot v - \frac{1}{n} \left( \Delta + n \right) T \, , \label{x2_u} \\[4pt]
\x_i & = r^2\, v_i + r\, D_i \x_r \, , \label{x2_i}
\end{align}
\end{subequations}
where $T$ and $v_i$ only depend on the angular coordinates $x^i$ and, when $n>2$, are constrained by the following differential conditions:
\begin{align}
D_{(i} v_{j)} - \frac{2}{n}\, \g_{ij} \Ddot v & = 0 \, , \label{kill2}\\[5pt]
D_{(i} D_{j)} T - \frac{2}{n}\, \g_{ij} \Delta T & = 0 \, . \label{killT2}
\end{align}
Eq.~\eqref{kill2}, which states that $v_i$ is a conformal Killing vector on the sphere at null infinity, actually holds in any space-time dimension, while $T$ has to satisfy \eqref{killT2} only when $n\neq 2$ (when $n=2$ the variations of the field components which are proportional to the combination \eqref{killT2} are of the same order as the falloffs \eqref{falloffs4D}). When $n > 2$, these conditions imply that the $\x_\m$ of the form \eqref{xi2} are Killing vectors of the Minkowski background, while for $n\leq 2$ (i.e.\ $D \leq 4$) a larger residual symmetry is allowed. When $n = 2$, the constraint \eqref{killT2} is indeed absent, so that supertranslations generated by an arbitrary $T(x^i)$ are allowed. Moreover, in this case, \eqref{kill2} admits locally infinitely many independent solutions, which generate superrotations. To analyse the $n = 1$ case, note that both \eqref{kill2} and \eqref{killT2} are traceless combinations: as a result, they vanish identically when $n=1$, so that both  $v_\phi(\phi)$ and $T(\phi)$ are arbitrary functions.

The asymptotically conserved charges corresponding to the previous residual symmetries are given by
\be \label{spin2_cov}
Q(u) = \lim_{r \to \infty} r^{n-1} \int_{S_u} d\Omega_n \left\{ h_{uu} \left( r\pr_r + n \right) \x_r + \frac{1}{r}\, \g^{ij} \left[\, \x_i\, \pr_r\, h_{uj} - h_{ui}\, \pr_r\, \x_j - \x_r D_i h_{uj} \,\right] \right\} ,
\ee
where the parameters $\x_\m$ are understood to satisfy eqs.~\eqref{xi2} and \eqref{kill2} (together with \eqref{killT2} when $n > 2$). The integral is evaluated over the sphere $S_u$ at constant retarded time on null infinity as, e.g., in \eqref{color_u} (see Appendix~\ref{app:charges} for more details). When $n$ is even, thanks to the limit $r \to \infty$, for fields satisfying the ansatz \eqref{ansatz2} $Q(u)$ reduces to 
\begin{align}
Q(u) = & - \sum_{k\,=\,0}^{\frac{n-2}{2}} r^{\frac{n-2k}{2}} \!\!\int_{S_u}\!d\Omega_n \left\{ \frac{n+2k+2}{2}\, v^i U^{(k)}_i + \left( T + \frac{u}{n}\, \Ddot v \right)\! \left[ n B^{(k-1)} + \frac{n+2k-2}{2\,r^2}\, \Ddot U^{(k)} \right] \right\} \nn \\
& - \int_{S_u}\! d\Omega_n \left\{ 2n\, m_B \left( T + \frac{u}{n}\, \Ddot v \right) + (n+1)\, \g^{ij} v_i N_j \right\} . \label{charge2_all}
\end{align}
When $n$ is odd, the only difference is that the extremum of the sum becomes $\frac{n-3}{2}$, so that the following considerations apply verbatim also to this case. As we shall see, at the linearised level, the $u-$dependence of the charges will turn out to be fictitious when $n \neq 2$.  This is actually expected on general grounds whenever the charges are computed on exact Killing vectors of the background \cite{Barnich-Brandt}, as recalled in Appendix \ref{app:charges}.

The integrals in the first line must vanish in order to have finite asymptotic charges and this is indeed the case if one considers the relations imposed by the equations of motion. From {the form}~\eqref{solU2} {of the solutions}, after integration by parts, one finds that, for each value of $k$, the first term in {the charge formula} \eqref{charge2_all} contains a~contribution proportional to
\be \label{div-zero-kill}
C^{(k)}_{ij} D^{(i} v^{j)} = C^{(k)}_{ij} \left( D^{(i} v^{j)} - \frac{2}{n}\, \g^{ij} \Ddot v \right) ,
\ee
where the trace condition $\g^{ij} C^{(k)}_{ij} = 0$ was also used.
Similarly, by {the relation} \eqref{solB2}, the second term in \eqref{charge2_all}, absent~when $n=2$, gives contributions than can be cast in the form
\be \label{totdiv2}
C^{(k)}_{ij} D^{(i} D^{j)}\!\left( T + \frac{u}{n}\, \Ddot v \right) = C^{(k)}_{ij} \left( D^{(i} D^{j)} T - \frac{2}{n}\, \g^{ij} \Delta T \right) + \frac{u}{n}\, C^{(k)}_{ij} D^{(i} D^{j)} \Ddot v \, .
\ee
This implies that {the combination} \eqref{div-zero-kill} vanishes in any space-time dimension (including $n=2$). The contribution in $T$ in \eqref{totdiv2} vanishes instead only when $n > 2$, but this does not cause any problem since the whole expression is actually absent when $n=2$. The last term in \eqref{totdiv2} vanishes as well when $n > 2$, since the divergence of the conformal Killing equation \eqref{kill2} implies
\be \label{identity_1}
\Delta v_i = \frac{2-n}{n}\, D_i \Ddot v - (n-1)\,v_i \, .
\ee
Acting with the Laplacian operator on \eqref{kill2} and substituting this identity, one eventually obtains
\be \label{trick}
(n-2)\, D_{(i} D_{j)} \Ddot v + 2\, \g_{ij} \left( \Delta + 2 \right) \Ddot v = 0 \, ,
\ee
which implies $C^{(k)}_{ij} D^{(i} D^{j)} \Ddot v = 0$ for a traceless $C^{(k)}_{ij}$ (when $n > 2$).

Similar arguments allow one to prove that the integral terms in \eqref{int-const2} (which are  absent when $n$ is odd---cf.\ \eqref{int-const2-odd}) do not contribute as well to the charges when $n>2$. In evaluating the last line of {the charge formula} \eqref{charge2_all}, one eventually has to take into account the precise $u$ dependence of $N_i$ dictated by \eqref{uU2}. For $n \neq 2$, this gives 
\be \label{final2}
Q = - \int_{S^n} \!d\Omega_n\, \Big\{ 2n\, T \cM + (n+1)\, v^i \cN_i \Big\} \, ,
\ee
 where the dependence on $u$ in the gauge parameter and in the field precisely cancels. The  linearised Poincar\'e charges \eqref{final2} depend on the ``integration constants'' that specify the Coulomb-type branch of the solutions of the equations of motion.\footnote{The same expression for the charges holds also when the dimension of spacetime is equal to three, and it corresponds to the natural presentation of $Q$ that one obtains in the Chern--Simons formulation of three-dimensional gravity (see e.g.\ Section~4.2 of \cite{review3D}). The only difference is that, when $n=1$, $T$ and $v^i$ are arbitrary functions of the angular coordinate $\phi$ on the circle at null~infinity.} Each integration constant, in its turn, is conjugated in \eqref{final2} to one of generators of the asymptotic symmetries. Let us also notice that these asymptotic charges, due to their constancy in $u$, should correspond in particular to the charges that one can measure at spatial infinity. A dependence on the retarded time, reflecting the changes in the total energy of the system induced by the flux of energy carried by the gravitational radiation, is reinstated when considering interactions \cite{gravity_evenD_1,positivity,gravity_evenD_2,gravity_anyD,angular-momentum}.

 The four-dimensional ($n=2$) case requires instead a separate analysis. As we have seen, the linear divergence in $r$ that appears in {the charge}  \eqref{charge2_all} in this case vanishes on account of {the identity} \eqref{totdiv2} as for generic $n$. By substituting {the expansions} \eqref{int4D} into the second line of \eqref{charge2_all}, one obtains instead 
\be
Q(u) = - \int_{S_u} \!\!d\Omega_n \left\{ T \left( 4\,\cM +  \Ddot\Ddot C \right) + v^i \left( 3\, \cN_i - \frac{u}{2}\, D_i \Ddot\Ddot C + D_i\! \int_{-\infty}^u \!\!du' \Ddot\Ddot C \right) \right\} .
\ee
The terms that appear for generic $n$ are reproduced, cf.~\eqref{final2}, but there is a sharp difference with respect to $n=2$: the charges now depend also on the boundary data of the radiation part of the solution, and this brings back a dependence on the retarded time. The charges associated with ordinary Poincar\'e transformations, however, still take the same form as in \eqref{final2}. For a translation, \eqref{killT2} indeed holds also in four dimensions. Similarly, for a Lorentz transformation, one actually has $D_{(i}D_{j)}\Ddot v = 0$ also when $n = 2$. These conditions are instead not satisfied by the $T$ and $v^i$ generating supertranslations and superrotations,  respectively (see e.g.\ \cite{bms-charges}). In the latter case, in~particular, in~order for the charges to be well defined, one should impose in addition suitable boundary conditions on $D_{i} \Ddot\Ddot C$ at the past boundary of null infinity. The corresponding dependence on $C_{ij}$ in the asymptotic charges is instrumental in deriving Weinberg's soft graviton theorem from the Ward identities of the supertranslation symmetry \cite{Weinberg-BMS}.\footnote{While this work was under completion, analogous, $u$-dependent asymptotic charges, associated with infinite-dimensional asymptotic symmetries  in any even space-time dimension, were presented in \cite{Pate_Memory}. In addition, in this case, the arbitrary function on the sphere generating the residual symmetry is conjugated to the boundary data of the radiation branch.} 

\vspace{6pt}
%%%%%%%%%%%%%%%%%%%%%%%%%%%%%%%%%%%%%%%%%%%%%%%%%%%%%%%%%%%%%%%%%%%%%
\section{Spin 3}\label{sec:spin3}
%%%%%%%%%%%%%%%%%%%%%%%%%%%%%%%%%%%%%%%%%%%%%%%%%%%%%%%%%%%%%%%%%%%%%

We now consider a single spin-3 field on a Minkowski background and we prove that the boundary conditions at null infinity proposed in \cite{super} give finite higher-spin charges in any space-time dimension. We work in the linearised theory, postponing to future work an analysis of the possible effects of the known cubic vertices\ft{For recent works and extensive references, see \cite{Metsaev_BRST,JLT,BPS,Sleight_etal,Conde_etal,FLM}.} on the asymptotic charges and their canonical algebra. 

%%%%%%%%%%%%%%%%%%%%%%%%%%%%%%%%%%%%%%%%%%
\subsection{Boundary Conditions}
%%%%%%%%%%%%%%%%%%%%%%%%%%%%%%%%%%%%%%%%%%

Following \cite{super}, we bound the spin-three field to satisfy the ``Bondi-like gauge''
\be \label{bondi3}
\vf_{r\m\n} = 0 \, , \quad \g^{ij} \vf_{ij\m} = 0\, , 
\quad \Rightarrow \quad g^{\m\n} \vf_{\m\n\r} = 0 \, ,
\ee
which is a natural generalisation of the Bondi gauge \eqref{bondi2} that we used in the analysis of linearised gravity. As in the latter case, the conditions \eqref{bondi3} cannot be reached with an off-shell gauge fixing, but~the number of constraints is the same as in the transverse-traceless gauge. We therefore assume that they can be imposed in a neighbourhood of null infinity and consider the reduced Fronsdal~equations\ft{Actually eq.~\eqref{fronsdal}, as well as its counterpart \eqref{maxwell-like-s} for spin $s$, follows from the Lagrangian equations of an alternative formulation for massless particles of any spin \cite{lowspin, ML, connections}.}
\be\label{fronsdal}
\cF_{\m\n\r} = \Box \vf_{\m\n\r} - \nabla_{\!(\m} \nabla\cdot \vf_{\n\r)} = 0 \, .
\ee

The expansion of \eqref{fronsdal} in Bondi coordinates has been presented in Appendix~A of \cite{super}, and can be extracted from the general spin-$s$ expressions presented below in \eqref{Frus} and \eqref{Fus}. 
Following the same logic as in the previous section, one can substitute a power-law ansatz in the equations of motion and realise that they are satisfied at the leading order provided that 
\be \label{boundary3}
\vf_{uuu} = \cO(r^{-\frac{n}{2}}) \, , \quad
\vf_{uui} = \cO(r^{-\frac{n}{2}+1}) \, , \quad
\vf_{uij} = \cO(r^{-\frac{n}{2}+2}) \, , \quad
\vf_{ijk} = \cO(r^{-\frac{n}{2}+3}) \, .
\ee

These are the fall-off conditions that have been proposed in \cite{super}; in the following, we shall exhibit the full solution of the linear equations of motion with these falloffs and we prove that it carries finite and non-trivial conserved spin-three charges at null infinity. As for gravity, the charges actually depend on the subleading terms at order $r^{1-n}$. We shall analyse separately even and odd space-time dimensions also in this case, while discussing in a dedicated subsection the main peculiarities emerging in three and four dimensions.

%%%%%%%%%%%%%%%%%%%%%%%%%%%%%%%%%%%%%%%%%%
\subsubsection{Even Space-Time Dimension}
%%%%%%%%%%%%%%%%%%%%%%%%%%%%%%%%%%%%%%%%%%

When $n$ is even, we then consider the following ansatz for the fields in the Bondi gauge \eqref{bondi3}:
\begin{subequations} \label{ansatz3}
\begin{alignat}{5}
\vf_{uuu} & = \sum_{l\,=\,0}^\infty r^{-\frac{n}{2}-l} B^{(l)}(u,x^m) \, , \qquad & 
\vf_{uui} & = \sum_{l\,=\,0}^\infty r^{-\frac{n}{2}-l+1} U^{(l)}_i(u,x^m) \, , \\
\vf_{uij} & = \sum_{l\,=\,0}^\infty r^{-\frac{n}{2}-l+2} V^{(l)}_{ij}(u,x^m) \, , \qquad &
\vf_{ijk} & = \sum_{l\,=\,0}^\infty r^{-\frac{n}{2}-l+3} C^{(l)}_{ijk}(u,x^m)  \, ,
\end{alignat}
\end{subequations}
with $\g^{ij}V_{ij}^{(l)} = \g^{ij} C_{ijk}^{(l)} = 0$.
This choice is further motivated by the observation that, in complete analogy with the lower-spin cases, a solution of this type gives rise to a generically non-zero energy flux through null infinity:
\be \label{energy-flux3}
\cP(u) = \lim_{r\to \infty} \int_{S_u} \left(T_{uu}-T_{ur}\right) r^n d\Omega_n =
	\int_{S_u} \gamma^{i_1j_1}\gamma^{i_2 j_2}\gamma^{i_3 j_3}\,\dot{C}^{(0)}_{i_1i_2i_3} \dot{C}^{(0)}_{j_1 j_2j_3} d\Omega_n \, ,
\ee
where $T_{\m\n}$ denotes the energy-momentum tensor of the solution (see Section~\ref{sec:spin-s} for more details).
It can be therefore interpreted as a spin-three wave reaching null infinity.

Substituting the ansatz \eqref{ansatz3} in the equations $\cF_{r\m\n} = 0$, one obtains
\begin{subequations} \label{sol3}
\begin{alignat}{5}
B^{(k)} & = \frac{2(n+2k-2)}{(n+2k)(n-2k-2)}\, \Ddot U^{(k)} & \quad \textrm{for}\ k & \neq \frac{n-2}{2} \, , \label{solB3}\\[10pt]
U^{(k)}_i & = \frac{2(n+2k)}{(n+2k+2)(n-2k)}\, \Ddot V^{(k)}_i \quad & \textrm{for}\ k & \neq \frac{n}{2} \, , \label{solU3}\\[10pt]
V^{(k)}_{ij} & = \frac{2(n+2k+2)}{(n+2k+4)(n-2k+2)}\, \Ddot C^{(k)}_{ij} \quad & \textrm{for}\ k & \neq \frac{n+2}{2} \, . \label{solV3}
\end{alignat}
\end{subequations}
In the cases excluded from the previous formulae, the coefficients in front of, respectively, $B^{(\frac{n-2}{2})}$, $U^{(\frac{n}{2})}$ and $V^{(\frac{n+2}{2})}$ vanish and the equations of motion imply instead
\be \label{constrC3}
(n-2)\, \Ddot\Ddot\Ddot C^{(\frac{n-2}{2})}_{\phantom{i}} = 0 \, , \qquad
\Ddot\Ddot C^{(\frac{n}{2})}_i = 0 \, , \qquad
\Ddot C^{(\frac{n+2}{2})}_{ij} = 0 \, .
\ee
The factor $(n-2)$ in the first constraint shows that $C^{(0)}$ remains arbitrary even when $n = 2$.
Substituting the same ansatz in the equation $\cF_{ijk} = 0,$ one obtains, for $l \neq \frac{n+2}{2}$,
\be \label{solC3}
\begin{split}
\dot{C}^{(l+1)}_{ijk} & = -\, \frac{1}{2(l+1)}\, \bigg\{ \left[\, \Delta - \frac{n(n-2)}{4} + l(l+1) - 3 \,\right] C^{(l)}_{ijk} \\
& - \frac{4}{(n+2l+4)(n-2l+2)} \left[\, (n+2)\, D_{(i}\Ddot C_{jk)}{}^{\!\!(l)} - 2\, \g_{(ij} \Ddot\Ddot C_{k)}{}^{\!\!(l)} \,\right] \bigg\} \, .
\end{split}
\ee
The value of $l$ excluded from this expression shows that from there on the tensors $C^{(l)}$ also depend on~$V^{(\frac{n+2}{2})}$,
\be \label{solC3-special}
\dot{C}^{(\frac{n+4}{2})}_{ijk} = \frac{1}{n+4}\, \bigg\{ D^{\phantom{(\frac{n}{2})}}_{(i} \!\!\!\! V^{(\frac{n+2}{2})}_{jk)} - \frac{2}{n+2}\, \g_{(ij}^{\phantom{(\frac{n}{2})}}\!\! \Ddot V_{k)}^{(\frac{n+2}{2})} - \left( \Delta + 2n - 1 \right) C^{(\frac{n+2}{2})}_{ijk} \bigg\} \, ,
\ee
although---as for gravity---these terms will not play any role in the analysis of the linearised charges.
The $u$-evolution of $B^{(\frac{n-2}{2})}$, $U^{(\frac{n}{2})}$ and $V^{(\frac{n+2}{2})}$ is fixed instead by the equations $\cF_{u\m\n} = 0$ (with $\m, \n \neq r$)~as
%\begingroup\makeatletter\def\f@size{9}\check@mathfonts
%\def\maketag@@@#1{\hbox{\m@th\normalsize\normalfont#1}}%
\begin{subequations} \label{u3}
\begin{align}
B^{(\frac{n-2}{2})} & = \cM - \frac{n-3}{6(n+1)n} \int_{-\infty}^u \!\!du'\!\left( \Delta - n + 2 \right) \Ddot\Ddot\Ddot C^{(\frac{n-4}{2})} ,\label{uB3} \\[10pt]
U_i^{(\frac{n}{2})} & = \cN_i + \frac{u}{n+2}\, \pr_i \cM - \frac{n-1}{2(n+1)(n+2)} \int_{-\infty}^u \!\!du'\! \left( \Delta - 1 \right) \Ddot\Ddot C_i^{(\frac{n-2}{2})} \nn \\
& - \frac{n-3}{6n(n+1)(n+2)} \int_{-\infty}^u \!\!du'\! \int_{-\infty}^{u'} \!\!du''\, D_i \left( \Delta - n + 2 \right) \Ddot\Ddot\Ddot C^{(\frac{n-4}{2})} \, , \label{uU3} \\[10pt]
V_{ij}^{(\frac{n+2}{2})} & = \cP_{ij} + \frac{u}{n+3} \left( D_{(i\,} \cN_{j)} - \frac{2}{n}\, \g_{ij} \Ddot \cN \right) + \frac{u^2}{2(n+2)(n+3)} \left( D_{(i} D_{j)} \cM - \frac{2}{n}\, \g_{ij}\, \Delta \cM \right) \nn \\
& - \frac{n+1}{(n+2)(n+3)} \int_{-\infty}^u \!\!du'\! \left( \Delta + n - 2 \right) \Ddot C_{ij}^{(\frac{n}{2})} + \cdots \, . \label{uV3}
\end{align}
\end{subequations}
%\endgroup
%
The omitted terms in the last equation correspond to the multiple integrals in the retarded time that one obtains by integrating the differential equation
\be \label{diff-eq-V}
\dot{V}_{ij}^{(\frac{n+2}{2})} = \frac{1}{n+3} \left\{ D_{(i}^{\phantom{(\frac{n}{2})}}\!\!\!\! U_{j)}^{(\frac{n}{2})} - \frac{2}{n}\, \g_{ij}^{\phantom{(\frac{n}{2})}}\!\!\!\! \Ddot U^{(\frac{n}{2})} - \frac{n+1}{n+2} \left( \Delta + n - 2 \right) \Ddot C_{ij}^{(\frac{n}{2})} \right\} 
\ee
given by the equation $\cF_{uij} = 0$. At any rate, in Section~\ref{sec:symm3}, we shall show that all integrals in these expressions do not contribute to the linearised charges, provided that one impose suitable regularity conditions that make them finite. 

The relevant terms to determine the charges are therefore those depending on the ``integration constants'' $\cM$, $\cN_i$ and $\cP_{ij}$, which actually admit an arbitrary dependence on the coordinates $x^m$ on the sphere at null infinity. They all appear at order $r^{n-1}$ in the expansions \eqref{ansatz3} and enter \eqref{u3} in combinations with a fixed polynomial dependence on the retarded time $u$.
For all other powers of $1/r$, the equations $\cF_{u\m\n} = 0$ (with $\m, \n \neq r$) reduce to divergences of \eqref{solC3} and are therefore identically satisfied (see \eqref{expansion-Fu} below for more details). As in the examples with lower spin, the divergences of \eqref{solC3} also imply the constraints \eqref{constrC3} (to be more precise their derivative in $u$). As a result, the~latter do not impose any further condition on the $C^{(l)}$ with lower values of $l$. Let us stress that some of the considerations above are valid only for $n > 2$; see Section~\ref{sec:3-4dim_3} for a discussion of the four- and three-dimensional cases.

%%%%%%%%%%%%%%%%%%%%%%%%%%%%%%%%%%%%%%%%%%
\subsubsection{Odd Space-Time Dimension}
%%%%%%%%%%%%%%%%%%%%%%%%%%%%%%%%%%%%%%%%%%

In complete analogy with linearised gravity, when $n$ is odd, one has to add further terms to the ansatz \eqref{ansatz3} in order to obtain non-trivial asymptotic charges at null infinity. We therefore consider the~ansatz
\begin{subequations} \label{ansatz3-odd}
\begin{alignat}{5}
\vf_{uuu} & = \vf_{uuu}[B] + \sum_{l\,=\,0}^\infty r^{1-n-l} \tilde{B}^{(l)}(u,x^m) \, , \qquad & 
\vf_{uui} & = \vf_{uui}[U] + \sum_{l\,=\,0}^\infty r^{1-n-l} \tilde{U}^{(l)}_i(u,x^m) \, , \\
\vf_{uij} & = \vf_{uij}[V] + \sum_{l\,=\,0}^\infty r^{1-n-l} \tilde{V}^{(l)}_{ij}(u,x^m) \, , \qquad &
\vf_{ijk} & = \vf_{ijk}[C] + \sum_{l\,=\,0}^\infty r^{1-n-l} \tilde{C}^{(l)}_{ijk}(u,x^m)  \, ,
\end{alignat}
\end{subequations}
where $\vf_{uuu}[B],$ etc.\ denote the terms introduced in {the expansions} \eqref{ansatz3}, which are still necessary if one desires to describe radiation, that is if one wishes to have a non-vanishing energy flux through null infinity (which is still given by \eqref{energy-flux3}). The new contributions to the expansion of the field components satisfy $\g^{ij} \tilde{V}_{ij}^{(l)} = \g^{ij} \tilde{C}_{ijk}^{(l)} = 0$.
Since $n$ is odd, all factors entering the expansion of the equations of motion in powers of $\sqrt{r}$ are different from zero. As a result, the tensors $B^{(l)}$, $U^{(l)}$ and $V^{(l)}$ satisfy the same conditions as in \eqref{sol3}, but without any constraint on the allowed values of $l$. Similarly, the tensors $C^{(l)}$ satisfy \eqref{solC3} for any $l$. The tensors appearing at the leading order of the new, Coulomb-like branch of our ansatz must satisfy
\begin{subequations} \label{u3-odd}
\begin{align}
\tilde{B}^{(0)} & = \cM \, , \qquad 
\tilde{U}_i^{(0)} = \cN_i + \frac{u}{n+2}\, \pr_i \cM \, ,  \\[10pt]
\tilde{V}_{ij}^{(0)} & = \cP_{ij} + \frac{u}{n+3} \left( D_{(i\,} \cN_{j)} - \frac{2}{n}\, \g_{ij} \Ddot \cN \right) + \frac{u^2}{2(n+2)(n+3)} \left( D_{(i} D_{j)} \cM - \frac{2}{n}\, \g_{ij}\, \Delta \cM \right) , 
\end{align}
\end{subequations}
on account of the equations of motion $\cF_{u\m\n} = 0$ (with $\m, \n \neq r$). Notice the similarity with eqs.~\eqref{u3}: the only difference is that, when $n$ is odd, there is no contribution from the data of the solution that encode radiation (here stored in the $\sqrt{r}$ branch). As we shall see below, the latter terms anyway do not contribute to the linearised charges. The subleading $\cO(r^{-n})$ terms in our ansatz do not contribute as well to the linearised charges. For this reason, we refrain from displaying here the relations that they have to satisfy in order to solve the equations of motion. The interested reader can extract them from the general expression \eqref{gen-Ctilde} presented in Section~\ref{sec:spin-s}.

%%%%%%%%%%%%%%%%%%%%%%%%%%%%%%%%%%%%%%%%%%
\subsubsection{Three and Four Space-Time Dimensions}\label{sec:3-4dim_3}
%%%%%%%%%%%%%%%%%%%%%%%%%%%%%%%%%%%%%%%%%%

When $n = 1$ or $n = 2$, the previous analysis has to be modified, in complete analogy with what we have seen for spin two. We begin by revisiting the four-dimensional case, where, in the component $\vf_{uuu}$, the leading order of the radiation and of the Coulomb-like solutions now coincide. Moreover, the~equation $\cF_{ruu} = 0$ does not impose any constraint on the triple divergence of $C^{(0)}$. As a result, the~equations fixing the dependence on $u$ of $B^{(0)}$, $U^{(1)}$ and $V^{(2)}$ are slightly modified as follows:
\begin{subequations} \label{int4D-spin3}
\begin{align}
B^{(0)} & = \cM + \frac{1}{6}\, \Ddot\Ddot\Ddot C^{(0)} \, , \\[10pt]
U_i{}^{(1)} & = \cN_i + \frac{u}{4}\, \pr_i \cM - \frac{1}{24} \int_{-\infty}^u \!\!du'\! \left[ \left( \Delta - 1 \right) \Ddot\Ddot C_i{}^{(0)} - 2\, D_i \Ddot\Ddot\Ddot C^{(0)} \right] , \\[10pt]
V_{ij}{}^{(2)} & = \cP_{ij} + \frac{u}{5} \left( D_{(i\,} \cN_{j)} - \g_{ij} \Ddot \cN \right) + \frac{u^2}{40} \left( D_{(i} D_{j)} \cM -  \g_{ij}\, \Delta \cM \right) + \cdots \, . \label{V4D}
\end{align}
\end{subequations}
$\cM$, $\cN_i$ and $\cP_{ij}$ are arbitrary functions of $x^i$ as in \eqref{u3} while in \eqref{V4D} we omitted the integrals that are obtained by the substitution of the previous formulae in the differential equation \eqref{diff-eq-V}, which is not modified even when $n =2$. In analogy with the spin-two case, the additional terms in \mbox{$\Ddot\Ddot\Ddot C^{(0)}$} will be instrumental in building the charges associated with the spin-three generalisations of supertranslations and superrotations identified in \cite{super}.

 When $n = 1$, the radiation branch becomes again subleading with respect to the Coulomb-type one in $\vf_{uuu}$. Moreover, fields of  spin greater than one do not propagate local degrees of freedom in three dimensions. It is therefore natural to ignore the radiation branch and work with boundary conditions that only encompass Coulomb-type solutions of the equations of motion. The only non-vanishing components of the field in the Bondi gauge \eqref{bondi3} are
\be 
\vf_{uuu} = \cM(\phi) + \cO(r^{-1}) \, , \qquad
\vf_{uu\phi} = \cN(\phi) + \frac{u}{3}\, \pr_\phi \cM(\phi) +\cO(r^{-1}) \, ,
\ee
where $\phi$ denotes again the angular coordinate on the circle at null infinity and we already included the constraints on the leading terms imposed by the equations of motion. The same conditions on the field $\vf_{\m\n\r}$ have been previously obtained in \cite{spin3-3D_1,spin3-3D_2} by translating boundary conditions proposed in the Chern--Simons formulation of three-dimensional spin-three gravity. The latter were designed to obtain asymptotic symmetries given by a contraction of the $\cW_3 \oplus \cW_3$ algebra of asymptotic symmetries of spin-three gravity in $AdS_3$ \cite{W3}. In analogy with the metric-like analysis performed in $AdS_3$ in \cite{metric3D}, in~the following we will recover the same infinite dimensional symmetries within our setup.

%%%%%%%%%%%%%%%%%%%%%%%%%%%%%%%%%%%%%%%%%%
\subsection{Asymptotic Symmetries and Charges}\label{sec:symm3}
%%%%%%%%%%%%%%%%%%%%%%%%%%%%%%%%%%%%%%%%%%

We now recall the key features of gauge transformations preserving the fall-off conditions \eqref{boundary3}, which have been identified in \cite{super}. Our current goal is to prove that the linearised charges associated with these asymptotic symmetries are finite. In Appendix~\ref{app:charges}, we show that, in the Bondi gauge \eqref{bondi3}, the~charges are expressed in terms of the fields and the parameters of asymptotic symmetries as
\be \label{spin3_cov}
\begin{split}
Q(u) & = -\lim_{r \to \infty} \frac{r^{n-1}}{2} \int_{S_u} d\Omega_n\, 
\bigg\{\,  r\,\vf_{uuu} \,\pr_r\, \x_{rr} + \x_{rr} \left( r\pr_r + 2n \right) \vf_{uuu} - \frac{2}{r}\, \x_{rr} \Ddot \vf_{uu} \\
& - \frac{2}{r^2}\, \g^{ij} \left[ \vf_{uui} \left( r\pr_r + n \right) \x_{rj} - \frac{1}{r}\, \x_{ri} \Ddot \vf_{uj} \right] + \frac{1}{r^3}\, \g^{ik}\g^{jl} \left[\, \vf_{uij} \,\pr_r\, \x_{kl} - \x_{ij} \,\pr_r\,  \vf_{ukl} \,\right] \bigg\} \, .
\end{split}
\ee

As we shall see, also in this case at the linearised level the $u-$dependence of the charges will turn out to be fictitious when $n \neq 2$.
(For $n > 2$, this is due to the fact that the only symmetries are exact symmetries of the background.) As far as the computation of charges is concerned, the only relevant components of the gauge parameters generating the residual gauge symmetry are
\begin{subequations} \label{asymptotic3}
\begin{gather}
\hspace{-55pt}\x_{rr}  = T - \frac{2u}{n+1}\, \Ddot \r + \frac{u^2}{(n+1)(n+2)}\, \Ddot\Ddot K \, , \\[5pt]
\hspace{-100pt}\x_{ri}  = r^2 \left( \r_i - \frac{u}{n+2}\, \Ddot K_i \right) + \frac{r}{2}\, D_i \x_{rr} \, , \\[5pt]
\begin{aligned}
\x_{ij} & = r^4\, K_{ij} + r \left( D_{(i} \x_{j)r} - \frac{2}{n+1}\, \g_{ij}\, \Ddot \x_r \right)  \\
 &- \frac{r^2}{4} \left( D_{(i} D_{j)} \x_{rr} - \frac{2(n+3)}{(n+1)(n+2)}\, \g_{ij} \left( \Delta + \frac{2(n+1)}{n+3} \right) \x_{rr} \right) .
\end{aligned}
\end{gather}
\end{subequations}
Here, $K_{ij}$, $\r_i$ and $T$ are tensors defined on the sphere at null infinity, which generalise the vector $v_i$ and the function $T$ that characterised the asymptotic symmetries of linearised gravity in \eqref{xi2}. They~thus depend only on the coordinates $x^m$, while all dependence on $u$ is explicit in the expressions above, which have been determined assuming an expansion in powers of $r$ and $u$.
The remaining components $\x_{uu}$, $\xi_{ur}$ and $\xi_{ui}$ are also non-vanishing and depend on $K_{ij}$, $\r_i$ and $T$. We refer to \cite{super} for their explicit expressions.

 The tensors $K_{ij}$, $\r_i$ and $T$ that characterise the asymptotic symmetries are not arbitrary: for $n > 2$, they are bound to satisfy the  differential equations
\begin{align}
\cK_{ijk} & \equiv D_{(i} K_{jk)} - \frac{2}{n+2}\, \g_{(ij} \Ddot K_{k)} = 0 \, , \qquad \g^{ij} K_{ij} = 0 \, , \label{K} \\[5pt]
\cR_{ijk} & \equiv D_{(i} D_j \r_{k)} - \frac{2}{n+2} \left(\, \g_{(ij} \Delta \r_{k)} + \g_{(ij} \big\{ D_{k)} , D_l \big\} \r^l \,\right) = 0 \, , \label{R} \\[5pt]
\cT_{ijk} & \equiv D_{(i} D_j D_{k)} T - \frac{2}{n+2} \left(\, \g_{(ij} \Delta D_{k)} T + \g_{(ij} \big\{ D_{k)} , D_l \big\} D^l T \,\right) = 0 \, . \label{T}
\end{align}

When the dimension of spacetime is equal to four, i.e.,\ when $n=2$, the last condition does not apply and the function $T(x^m)$ is instead arbitrary \cite{super}. In this case, the corresponding symmetry is the analogue of gravitational supertranslations. Eq.~\eqref{K} generalises the conformal Killing equation~\eqref{kill2} and states that $K_{ij}$ is a conformal (traceless) Killing tensor of rank-two on the celestial sphere (see e.g.\ \cite{conformal-killing}). For $n > 2$, this equation admits a finite number of solutions, while, when $n = 2$, locally there are infinitely many solutions, generalising gravitational superrotations. The same is true for the less familiar equation~\eqref{R} satisfied by $\r_i$: when $n>2$, it admits a finite number of solutions, while, for $n = 2$,
locally one can build infinitely many independent solutions. For the details, we refer again to~\cite{super}.  All combinations above are traceless: as a result, they vanish identically when the dimension of spacetime is equal to three, i.e.,\ when $n = 1$. This implies that, in three dimensions, $T(\phi)$ and $\r(\phi)$ are arbitrary functions, while the symmetry generated by the traceless $K_{ij}$ is actually absent.
 
Substituting {the gauge parameters}~\eqref{asymptotic3} into the expression for the charges \eqref{spin3_cov}, one obtains
\be \label{charge3-mid}
\begin{split}
Q(u) = \lim_{r \to \infty} \frac{r^{n-1}}{2} \int_{S_u} \!d\Omega_n \, \bigg\{\, \chi & \left( T - \frac{2u}{n+1}\, \Ddot \r + \frac{u^2}{(n+1)(n+2)}\, \Ddot\Ddot K \right) \\
+\, \chi^i & \left( \r_i - \frac{u}{n+2}\, \Ddot K_i \right) + \chi^{ij} K_{ij} \,\bigg\} \, ,
\end{split}
\ee
with
\begin{subequations}
\begin{align}
\chi & = - \left( r\pr_r + 2n \right) \vf_{uuu} - \frac{n-1}{r}\, \Ddot \vf_{uu} + \frac{1}{2r}\, \pr_r \Ddot\Ddot \vf_u \, , \\
\chi_i & = 2(n+2)\,\vf_{uui} - \frac{2}{r} \left( r\pr_r - 2 \right) \Ddot \vf_{ui} \, , \\[3pt]
\chi_{ij} & = \left( r\pr_r -4 \right) \vf_{uij} \, .
\end{align}
\end{subequations}

The next task is to evaluate {the charge} \eqref{charge3-mid} on the solutions of the equations of motion discussed above. For~$n$ even and greater than two \eqref{sol3}, \eqref{constrC3} and \eqref{u3} lead to
\begingroup\allowdisplaybreaks
\makeatletter\def\f@size{9}\check@mathfonts
\def\maketag@@@#1{\hbox{\m@th\normalsize\normalfont#1}}%
\begin{subequations} \label{all-chi}
\begin{align}
\chi & = - \sum_{k\,=\,0}^{\frac{n-4}{2}} r^{-\frac{n}{2}-k} \frac{n+2k-2}{2(n-2k-2)}\, \Ddot\Ddot\Ddot C^{(k)} - r^{1-n} (n+1) \cM \nn \\*
& + r^{1-n}\, \frac{n-3}{6n} \int_{-\infty}^u \!\!du'\!\left( \Delta - n + 2 \right) \Ddot\Ddot\Ddot C^{(\frac{n-4}{2})} + o(r^{1-n}) \, , \label{chi} \\[5pt]
\chi_i & = \sum_{k\,=\,0}^{\frac{n-2}{2}} r^{-\frac{n}{2}-k+1}\, \frac{2(n+2k)}{n-2k}\, \Ddot\Ddot C_i{}^{\!(k)} + 2\, r^{1-n} \Big\{ (n+2)\, \cN_i + u\, \pr_i \cM \Big\} + o(r^{1-n}) \nn \\*
& - \frac{r^{1-n}}{n+1} \int_{-\infty}^u \!\!\!\!du'\! \left\{ (n-1) \left( \Delta - 1 \right) \Ddot\Ddot C_i{}^{\!(\frac{n-2}{2})} + \frac{n-3}{3n} \int_{-\infty}^{u'} \!\!\!\!du'' D_i\! \left( \Delta - n + 2 \right)(\Ddot)^3 C^{(\frac{n-4}{2})} \right\} , \label{chi1} \\[5pt]
\chi_{ij} & = - \sum_{k\,=\,0}^{\frac{n}{2}} r^{-\frac{n}{2}-k+2} \frac{n+2k+2}{n-2k+2}\, \Ddot C_{ij}{}^{\!(k)} \label{chi2} \\*
& - r^{1-n} \left\{ (n+3)\, \cP_{ij} + u \left( D_{(i\,} \cN_{j)} - \frac{2}{n}\, \g_{ij} \Ddot \cN \right) + \frac{u^2}{2(n+2)} \left( D_{(i} D_{j)} \cM - \frac{2}{n}\, \g_{ij}\,\Delta \cM \right) \right\} + \cdots \, , \nn
\end{align}
\end{subequations}
\endgroup
where, in \eqref{chi2}, besides the terms $o(r^{1-n})$, we also omitted the integrals in the retarded time that one~obtains by substituting \eqref{uV3}.  For \mbox{$n=2$}, the first two expressions are modified as follows:
\begingroup\makeatletter\def\f@size{9}\check@mathfonts
\def\maketag@@@#1{\hbox{\m@th\normalsize\normalfont#1}}%
\begin{subequations} \label{chi4D}
\begin{align}
\chi & = - \frac{1}{r} \left\{ 3\, \cM + \frac{1}{2}\, \Ddot\Ddot\Ddot C^{(0)} \right\} + \cO(r^{-2}) \, , \\[5pt]
\chi_i & = 2\,\Ddot\Ddot C_i{}^{\!(0)} + \frac{2}{r} \left\{ 4\, \cN_i + u\, \pr_i \cM + \frac{1}{6} \int_{-\infty}^u \!\!\!\!du'\! \left[ \left( \Delta - 1 \right) \Ddot\Ddot C_i{}^{\!(0)} - 2\,D_i \Ddot\Ddot\Ddot C^{(0)} \right] \right\} \nn \\
& + \cO(r^{-2}) \, .
\end{align}
\end{subequations}\endgroup
The correct $\chi_{ij}$ is instead obtained by setting $n = 2$ in \eqref{chi2} and by correcting the integral terms according to \eqref{V4D}. For $n$ odd, the extrema of the sums become, respectively, $\frac{n-3}{2}$, $\frac{n-1}{2}$ and $\frac{n+1}{2}$, while~the terms in the second lines of eqs.~\eqref{all-chi} are absent.

 Looking only at the $r$-dependence, the sums in the previous formulae would give a divergent contribution to the charges. These vanish, however, thanks to the differential constraints on the parameters in \eqref{K}, \eqref{R} and, when, $n > 2$, \eqref{T}. Let us begin by exhibiting this mechanism in the simplest case: the term $\chi^{ij} K_{ij}$ in {the charge} \eqref{charge3-mid} contains divergent contributions that, integrating by parts, can be cast in the form
\be
C_{ijk}^{(l)}\, D^{(i} K^{jk)} = C_{ijk}^{(l)} \left\{ \frac{2}{n+2}\,\g^{(ij} \Ddot K^{k)} - \cK^{ijk} \right\} = 0 \, ,
\ee
where we recall that $\cK_{ijk}$ is the shortcut introduced in \eqref{K} to denote the differential equation satisfied by $K_{ij}$.
This cancellation is the analogue of the one involving the conformal Killing equation in linearised gravity: it holds because the conformal Killing tensor equation allows for substituting the symmetrised gradient with a term in $\g_{ij}$ and the tensors $C^{(l)}$ are traceless. The next cancellation is slightly more involved: integrating by parts one obtains
\be
\int_{S_u} \!\!\!d\Omega_n\ \chi^i \left( \r_i - \frac{u}{n+2}\, \Ddot K_i \right) \sim \sum_{l\,=\,0}^{[\frac{n-1}{2}]} \!r^{-\frac{n}{2}-l}\! \int_{S_u} \!\!\!d\Omega_n\  C_{ijk}^{(l)} \left( D^{(i} D^{j} \r^{k)} - \frac{u}{n+2}\, D^{(i} D^{j} \Ddot K^{k)} \right) + \cdots \, .
\ee
To cancel the contribution in $\r_i$, one can use eq.~\eqref{R}, which again allows one to substitute the symmetrised gradient with a term in $\g_{ij}$. To cancel the contribution in $K_{ij}$, one can instead use the following consequence of the conformal Killing tensor equation \eqref{K}: 
\be
\begin{split}
D_{(i} D_j \Ddot K_{k)} & = -\, 2\, \g_{(ij} \Ddot K_{k)} + \frac{3}{n+1}\, \g_{(ij} D_{k)} \Ddot\Ddot K \\
& - \frac{n+2}{n} \left\{ \Delta \cK_{ijk} - D_{(i} \Ddot \cK_{jk)} + \frac{1}{n+1}\, \g_{(ij} \Ddot\Ddot \cK_{k)} + (n-3)\, \cK_{ijk} \right\} .
\end{split}
\ee
Similar considerations apply to the integral terms in the second line of \eqref{chi1}.
The remaining contribution in the charge formula \eqref{charge3-mid} contains three addenda whose divergent parts can be cast in the following form by integrating by parts:
\be
C_{ijk}^{(l)} D^{(i} D^{j} D^{k)} T \, , \qquad
C_{ijk}^{(l)} D^{(i} D^{j} D^{k)} \Ddot \r \, , \qquad
C_{ijk}^{(l)} D^{(i} D^{j} D^{k)} \Ddot\Ddot K \, .
\ee
These terms are actually absent when $n=2$. For $n>2$, the first contribution vanishes thanks to {the differential constraint} \eqref{T}. The other two types of terms vanish thanks to the following consequences of the equations satisfied by $\r_i$ and $K_{ij}$:
\begin{subequations}
\begin{align}
D_{(i} D_j D_{k)} \Ddot \r & = \frac{2}{n+2} \, \g_{(ij} D_{k)} \left( 3\Delta + 2(n-1) \right) \Ddot\r + \textrm{terms in}\ \cR_{ijk} \, , \\ 
D_{(i} D_j D_{k)} \Ddot\Ddot K & = -\, 8\, \g_{(ij} D_{k)} \Ddot\Ddot K + \textrm{terms in}\ \cK_{ijk} \, .
\end{align}
\end{subequations}
The precise form of the omitted terms is displayed in eqs.~(B.8) and (B.9) of \cite{super}.

 We have therefore proven that, in the Bondi gauge \eqref{bondi3}, a spin-three field with the falloffs \eqref{boundary3} at null infinity given in \cite{super} and satisfying the Fronsdal equations up to the contributions of order $r^{n-1}$ to its components admits finite asymptotic linearised charges. For any $n > 2$, these depend on the ``integration constants'' specifying the solution as
\be \label{finalcharge3}
Q = - \frac{1}{2} \int_{S^n} \!d\Omega_n\, \Big\{ (n+1)\, T \cM - 2(n+2)\, \r^i \cN_i + (n+3)\, K^{ij} \cP_{ij} \Big\} \, ,
\ee 
where $\cM$, $\cN_i$ and $\cP_{ij}$ are the tensors on the sphere at null infinity introduced in \eqref{u3} (cf. \eqref{u3-odd} for $n$~odd). As anticipated, the charges are constant along null infinity when the dimension of spacetime is greater than four. The same is true also in three dimensions: in this case, both $K^{ij}$ and $\cP_{ij}$ are not present and the asymptotic charges take the form
\be
Q = - \int d\phi \Big\{ T(\phi) \cM(\phi) - 3\, \r(\phi) \cN(\phi) \Big\}\, ,
\ee
in agreement with the result derived in the Chern--Simons formulation \cite{spin3-3D_1,spin3-3D_2}. When the dimension of spacetime is equal to three or bigger than four, with our boundary conditions, the spin-three charges thus display a structure very similar to that of the corresponding charges computed on anti de Sitter backgrounds in \cite{charges-AdS}. The latter, indeed, in the limit of vanishing cosmological constant should reproduce the flat-space charges at spatial infinity.

 When $n = 2$, the modifications in the dependence on $u$ of the leading terms in the Coulomb-type solution recalled in \eqref{int4D-spin3} (and \eqref{chi4D}) lead to the following expression for the asymptotic charges:
\be \label{charge3-4D}
Q(u) = - \frac{1}{2} \int_{S_u} \!\!d\Omega_n \left\{ 3\, T \left( \cM + \frac{1}{6}\, \Ddot\Ddot\Ddot C^{(0)} \right) - 8\, \r^i \cN_i + 5\, K^{ij} \cP_{ij} + \cdots \right\} .
\ee
In this formula, we omitted other $u$--dependent terms in $C^{(0)}$, whose form is not particularly illuminating and can be readily obtained by substituting \eqref{chi4D} in \eqref{charge3-mid}. The main information is anyway that in four dimensions a dependence on the retarded time appears already in the linearised theory, thanks to the contribution to the charges of the radiation solution. As shown in \cite{super}, where~the terms in $T$ in {the charge} \eqref{charge3-4D} have been actually already presented, the dependence on radiation data is instrumental in deriving Weinberg's theorem for spin-three soft quanta from the Ward identities of the supertranslation symmetry generated by the arbitrary function $T(x^i)$.

\vspace{6pt} 
%%%%%%%%%%%%%%%%%%%%%%%%%%%%%%%%%%%%%%%%%%%%%%%%%%%%%%%%%%%%%%%%%%%%%
\section{Arbitrary Spin}\label{sec:spin-s}
%%%%%%%%%%%%%%%%%%%%%%%%%%%%%%%%%%%%%%%%%%%%%%%%%%%%%%%%%%%%%%%%%%%%%

In this section, we first extend to arbitrary values of the spin the analysis of the linearised equations of motion displayed in Sections~\ref{sec:spin2} and \ref{sec:spin3} for fields of spin two and three. We then perform a preliminary study of the residual gauge symmetry preserving the form of the solutions. In particular, we fix the structure of asymptotic symmetries to leading order in an expansion in powers of $r$. This allows us both to exhibit some examples of on-shell cancellations of divergent contributions to the charges and to propose a general expression for their finite part in terms of the integration constants, arising after integration over $u$, which specify the solutions. This allows us to motivate a proposal  for boundary conditions giving finite asymptotic spin-$s$ charges in Minkowski backgrounds of any dimension. 

%%%%%%%%%%%%%%%%%%%%%%%%%%%%%%%%%%%%%%%%%%
\subsection{Boundary Conditions}
%%%%%%%%%%%%%%%%%%%%%%%%%%%%%%%%%%%%%%%%%%

In the retarded Bondi coordinates \eqref{bondi-coord}, we study Fronsdal's equations in the ``Bondi gauge'' defined~by
\be \label{Bondi-s}
\vf_{r\m_1 \cdots \m_{s-1}} = 0 \, , \qquad
\g^{ij} \vf_{ij \m_1 \cdots \m_{s-2}} = 0 \, .
\ee
These constraints imply that the fields are traceless, so that Fronsdal's tensor take the Maxwell-like form \cite{ML}
\be \label{maxwell-like-s}
\cF_{\m_s} = \Box \vf_{\m_{s}} - \nabla_{\!\m} \nabla\cdot \vf_{\m_{s-1}}\, .
\ee
Here and in the following, groups of symmetrised indices are denoted by a single Greek letter with a label indicating the total number of indices, so that, e.g.,\ $\vf_{\m_1 \cdots \m_s}\! \to \vf_{\m_s}$; repeated indices denote instead a symmetrisation involving the minimum number of terms needed and without any overall factor, so that, e.g.,\ $A_\m B_\m \equiv A_{\m_1} B_{\m_2} + A_{\m_2} B_{\m_1}$. For more details, see \cite{massive}.
As before, we also assume that their solutions can be expanded in (half-integer) powers of $r^{-1}$ in a neighbourhood of null infinity.

 Eq.~\eqref{Bondi-s} also implies that all components of the Fronsdal tensor with at least two radial indices vanish identically. The components with a single radial index read instead
\be \label{Frus}
\cF_{r\, u_{s-k-1} i_{k}} = \frac{1}{r^2} \left\{ r^2\pr_r^2 + \left( n-2k \right) r\pr_r - 2k(n-1) \right\} \vf_{u_{s-k} i_{k}} - \frac{1}{r^3} \left( r\pr_r - 2k \right) \Ddot \vf_{u_{s-k-1} i_{k}} \, .
\ee
The remaining components, without any radial index, are
\be \label{Fus}
\begin{split}
\cF_{u_{s-k} i_{k}} & = \frac{1}{r} \left\{ (s-k-2)\, r\pr_r + n(s-k-1) + 2k \right\} \dot{\vf}_{u_{s-k} i_{k}} - \frac{s-k}{r^2}\, \Ddot \dot{\vf}_{u_{s-k-1}i_{k}} \\
& + \frac{1}{r^2} \left\{ \left[ \Delta + r^2 \pr_r^2 + (n-2k)\, r\pr_r - k(n-k) \right] \vf_{u_{s-k} i_{k}} - D_i \Ddot \vf_{u_{s-k} i_{k-1}}\right\} \\
& + \frac{1}{r} \left\{ \left( r\pr_r + n -2 \right) D_i \vf_{u_{s-k+1}i_{k-1}} + 2\, \g_{ii} \Ddot \vf_{u_{s-k+1} i_{k-2}} \right\} - 2 \left( r\pr_r + n -1 \right) \g_{ii} \vf_{u_{s-k+2}i_{k-2}} \, .
\end{split}
\ee

 We begin by studying the previous equations for $n$ even. In this case, we employ the ansatz
\be \label{s-ansatz_even}
\vf_{u_{s-k}i_k} = \sum_{l\,=\,0}^\infty r^{-\frac{n}{2}+k-l} C_{i_k}{}^{\!\!\!(k,l)}(u,x^m) \, .
\ee
As in the previous examples, the leading behaviour of our ansatz is designed to give a finite flux of energy per unit time across the sphere $S_u$ at fixed $u$, a feature that we interpret as radiation crossing null infinity. The canonical energy-momentum tensor of the Fronsdal Lagrangian in Bondi gauge,
\be
\mathcal L = \frac{1}{2}\,  \varphi^{\m_s} \left(\, \Box \varphi_{\m_s} - \nabla_{\!\m} \nabla\cdot \varphi_{\m_{s-1}} + \eta_{\m\m} \nabla\cdot\nabla\cdot \varphi_{\m_{s-2}}\, \right) ,
\ee
reads indeed
\be
T_{\alpha\beta}= \nabla_{\!\alpha} \varphi_{\m_s} \nabla_{\!\beta} \varphi^{\m_s} - s\, \nabla\cdot \varphi^{\m_{s-1}} \nabla_{\!\alpha} \varphi_{\beta\m_{s-1}}+ \eta_{\a\b} \mathcal L \, .
\ee
The corresponding power flowing through null infinity is then 
\be \label{energy-flux}
\cP(u) =  \lim_{r\to \infty} \int_{S_u} \left(T_{uu}-T_{ur}\right) r^n d\Omega_n 
 = \int_{S_u} \gamma^{i_1 j_1} \cdots \gamma^{i_s j_s}\, \dot{C}^{(s,0)}_{i_1 \cdots i_s} \dot{C}^{(s,0)}_{j_1 \cdots j_s} d\Omega_n\,.
\ee

When substituting the ansatz \eqref{s-ansatz_even}, the components $\cF_{r\m_{s-1}}$ of the Fronsdal tensor vanish provided~that
\be \label{Ckl}
C_{i_k}{}^{\!\!\!(k,l)} = 
\frac{2\left[n+2(k+l-1)\right]}{\left[n+2(k+l)\right]\left[n+2(k-l-1)\right]}\, \Ddot C_{i_k}{}^{\!\!\!(k+1,l)} \qquad \textrm{for}\ l \neq \frac{n}{2}+ k -1 \, ,
\ee
while, for $l = n/2 + k -1$, the equations of motion imply
\be
(n+2k-2) \Ddot C_{i_k}{}^{\!\!\!(k+1,\frac{n}{2}+k-1)} = 0 \, .
\ee
In the formula above, we exhibited the factor emerging in the computation that vanishes when $n = 2$ and $k = 0$. This anticipates that the peculiarities of the four-dimensional case that we encountered before persist for arbitrary values of the spin. At any rate, \eqref{Ckl} shows that, for all even values of $n$, the~tensors entering the ansatz \eqref{s-ansatz_even} are fixed in terms of the $C^{(s,l)}$, with the exception of 
\be \label{def-charges}
C_{i_k}{}^{\!\!\!(k,\frac{n}{2}+k-1)} \equiv Q_{i_k}{}^{\!\!\!(k)} \qquad \textrm{for}\ k < s \, .
\ee

The tensors $C^{(s,l)}$ are then determined (up to integrations constants) in terms of an arbitrary tensor $C^{(s,0)}(u,x^m)$ via the equation $\cF_{i_s} = 0$, which gives
\be \label{Cs}
\begin{split}
\dot{C}_{i_s}{}^{\!\!\!(s,l+1)} & = - \frac{1}{2(l+1)} \bigg\{ \left[ \Delta - \frac{n(n-2)}{4} + l(l+1) - s \right] C_{i_s}{}^{\!\!\!(s,l)} \\
& - \frac{4(n+2s-4)}{[n+2(s+l-1)][n+2(s-l-2)]} \left[ D_i \Ddot C_{i_{s-1}}{}^{\!\!\!\!\!\!\!\!\!(s,l)} - \frac{2}{n+2s-4}\, \g_{ii} \Ddot\Ddot C_{i_{s-2}}{}^{\!\!\!\!\!\!\!\!\!(s,l)}  \right] \bigg\} \, .
\end{split}
\ee 

 The remaining components of the equations of motion fix the $u$-evolution of the tensors $\cQ^{(k)}$ defined in \eqref{def-charges}. To this end, it is convenient to expand the Fronsdal tensor in powers of $r^{-1}$. When {the ansatz} \eqref{s-ansatz_even} holds, one has
\be
\cF_{u_{s-k}i_k}[C] = \sum_{l\, =\, 0}^\infty r^{-\frac{n}{2}+k-l-1} \mathfrak{F}_{i_{k}}{}^{\!\!\!(k,l)}(u,x^m)
\ee 
and, when the equations $\cF_{r\m_{s-1}} = 0$ are satisfied, one can recast the expansion in the following form:
\be \label{expansion-Fu}
\begin{split}
\cF_{u_{s-k}i_k} & = \sum_{\substack{l\,=\,0 \\ l\,\neq\,\frac{n}{2}+k-2}}^\infty \frac{2\,r^{-\frac{n}{2}+k-l-2}}{n+2(k-l-2)}\, \Ddot \mathfrak{F}_{i_{k}}{}^{\!\!\!(k+1,l+1)} \\
& + r^{-n} \bigg\{ (n+s+k-2) \dot{\cQ}_{i_k}{}^{\!\!\!(k)} - D_i \cQ_{i_{k-1}}{}^{\!\!\!\!\!\!\!\!\!(k-1)} + \frac{2}{n+2k-4}\, \g_{ii} \Ddot \cQ_{i_{k-2}}{}^{\!\!\!\!\!\!\!\!\!(k-1)} \\
& + \frac{n+2k-3}{(s-k)!(n+s+k-3)} \left[ \Delta + k(k-4) + n(k-1) + 2 \right] (\Ddot)^{s-k} C^{(s,\frac{n}{2}+k-2)}\bigg\} \, .
\end{split}
\ee

This implies that, on shell, the $\cQ^{(k)}$ are $n$-divergences as the other $C^{(k,l)}$, however up to a set of integrations constants $\cM^{(k)}(x^m)$. The second line of \eqref{expansion-Fu} actually dictates that $\cQ^{(k)}$ depends on the integrations constants of all $\cQ^{(l)}$ with $l < k$ with a precise polynomial dependence on $u$. As we have seen in the previous examples, this is instrumental in making the independence on the retarded time of the asymptotic charges explicit. Concretely, the $Q^{(k)}$ depend on the integrations constants as 
\be \label{finalQ}
\cQ_{i_k}{}^{\!\!(k)} = \sum_{l\,=\,0}^k \frac{(n+s+k-l-2)!}{l!(n+s+k-2)!}\, u^l \underbrace{D_{i} \cdots D_i}_{\textrm{$l$ terms}} \cM_{i_{k-l}}{}^{\!\!\!\!\!\!\!\!(k-l)} + \cdots .
\ee
In this formula, we omitted both the terms in $\cM^{(k-l)}$ that make the expression above traceless and the terms in the $C^{(k,l)}$ resulting from the integration of \eqref{expansion-Fu}. Both types of contributions anyway will not contribute to the asymptotic charges when $n > 2$. When $n = 2$, eq.~\eqref{expansion-Fu} has to be modified in analogy with the discussions in Sections~\ref{sec:3-4dim_2} and \ref{sec:3-4dim_3}, since $(\Ddot)^{s} C^{(0)}$ does not vanish anymore.

When $n$ is odd, one has to consider an ansatz containing both integer and half-integer powers of $r$. For~$n > 1$, we set
\be \label{s-ansatz_odd}
\vf_{u_{s-k}i_k} = \sum_{l\,=\,0}^\infty r^{-\frac{n}{2}-l+k} C^{(k,l)}(u,x^m) + \sum_{l\,=\,0}^\infty r^{1-n-l} \tilde{C}^{(k,l)}(u,x^m) \, ,
\ee
so that the leading order has the same form as in {the ansatz} \eqref{s-ansatz_even}. Eq.~\eqref{energy-flux} thus guarantees that we have a~finite flux of energy per unit time across $S_u$ in this case too. When $n = 1$, consistently with the absence of propagating degrees of freedom for fields of spin $s>1$ in three dimensions, the radiation branch of the solution becomes subleading in the field component with only $u$ indices and we will ignore it as in the examples with spin two and three. Notice that, due to the trace constraint in \eqref{Bondi-s}, in~this case, the only non-vanishing components of the field are
\be
\vf_{u_s} = \cM^{(s)}(\phi) + \cO(r^{-1}) \, , \qquad
\vf_{u_{s-1}\phi} = \cN^{(s)}(\phi) + \frac{u}{s}\, \pr_\phi \cM^{(s)}(\phi) + \cO(r^{-1}) \, .
\ee
The boundary conditions therefore contain only two arbitrary functions for each value of the spin, in analogy with what happens in $AdS_3$ \cite{W3,Wextra1,Wextra2,Wextra3,metric3D}.

The equations $\cF_{r\m_{s-1}}= 0$ imply {the relations} \eqref{Ckl} also when $n$ is odd and greater than one, but without any limitation on the allowed values of $k$. The leading order of the Coulomb branch is instead not constrained by these equations, so that we can define
\be
\tilde{C}^{(k,0)} \equiv \cQ^{(k)} \qquad \textrm{for}\ k < s \, .
\ee

The remaining $\tilde{C}^{(k,l)}$ are again fixed in terms of the $\tilde{C}^{(s,l)}$ by \eqref{Ckl}, provided that one identifies $\tilde{C}^{(k,l)} = C^{(k,\frac{n}{2}+k+l-1)}$, which is
\be \label{gen-Ctilde}
\tilde{C}^{(k,l+1)} = - \frac{n+2k+l-1}{(l+1)(n+2k+l)}\, \Ddot \tilde{C}^{(k+1,l)} \, .
\ee
Performing the same substitution in \eqref{Cs} gives the relation fixing all $\tilde{C}^{(s,l)}$ in terms of an arbitrary $\tilde{C}^{(s,0)}(u,x^m)$ (again up to integration constants).

 Expanding the contributions to the Fronsdal tensor of the terms with integer powers of $r$ as
\be
\cF_{u_{s-k}i_k}[\tilde{C}] = \sum_{l = 0}^\infty r^{-n-l} \tilde{\mathfrak{F}}^{(k,l)}(u,x^m)\, ,
\ee
one eventually obtains
\be
\begin{split}
\cF_{u_{s-k}i_k} & = \sum_{l\,=\,0}^\infty \frac{2\,r^{-\frac{n}{2}+k-l-2}}{n+2(k-l-2)}\, \Ddot \mathfrak{F}_{i_{k}}{}^{\!\!\!(k+1,l+1)} - \sum_{l\,=\,0}^\infty \frac{r^{-n-l-1}}{l+1}\, \Ddot \tilde{\mathfrak{F}}_{i_{k}}{}^{\!\!\!(k+1,l)} \\
& + r^{-n} \left\{ (n+s+k-2) \dot{\cQ}_{i_k}{}^{\!\!\!(k)} - D_i \cQ_{i_{k-1}}{}^{\!\!\!\!\!\!\!\!\!(k-1)} + \frac{2}{n+2k-4}\, \g_{ii} \Ddot \cQ_{i_{k-2}}{}^{\!\!\!\!\!\!\!\!\!(k-1)} \right\} .
\end{split}
\ee
As a result, when $n$ is odd, the $\cQ^{(k)}$ satisfy a relation analogous to \eqref{finalQ}, where the omitted terms, which anyway do not contribute to the charges, are actually absent.

 By analogy with the examples of spin two and three, one is led to conclude that, in the Bondi gauge \eqref{Bondi-s}, the boundary conditions to be imposed on a spin-$s$ field in order to obtain finite asymptotic charges for any $D$, even~or odd, are the following:
\be \label{boundary-cond-s}
\begin{split}
\vf_{u_{s-k}i_k} = & \sum_{l\,=\,0}^{\left[\frac{n+1}{2}\right]+k-2} r^{-\frac{n}{2}+k-l}\, \frac{2^{s-k}(n+2(k-l-2))!!(n+2(k+l-1))}{(n+2(s-l-2))!!(n+2(s+l-1))}\, (\Ddot)^{s-k} C_{i_k}{}^{\!\!\!(s,l)} \\
& + r^{1-n} \cQ_{i_k}{}^{\!\!\!(k)} + \cO(r^{-\frac{n}{2}-\left[\frac{n}{2}\right]}) \, ,
\end{split}
\ee
where the $C^{(k,l)}$ satisfy \eqref{Ckl}, while the $\cQ^{(k)}$ satisfy \eqref{finalQ}.
In this work, we do not perform a complete analysis of the asymptotic symmetries for fields of arbitrary spin. Still, we can provide support to the correctness of {the boundary conditions} \eqref{boundary-cond-s} by showing that {they lead} to the cancellation of some of the potentially divergent contributions to the linearised charges, while also proving that the $\cQ^{(k)}$ give a finite contribution to them. This is the goal of the next section.

%%%%%%%%%%%%%%%%%%%%%%%%%%%%%%%%%%%%%%%%%%
\subsection{Asymptotic Symmetries and Charges}
%%%%%%%%%%%%%%%%%%%%%%%%%%%%%%%%%%%%%%%%%%

In order to preserve the boundary conditions \eqref{boundary-cond-s}, the variations of the field components in Bondi coordinates,
\be \label{variation-s}
\begin{split}
\d \vf_{r_{s-k-l}u_li_k} & = l\, \dot{\x}_{r_{s-k-l}u_{l-1}i_k} + \frac{s-k-l}{r} \left( r\pr_r - 2k \right) \x_{r_{s-k-l-1}u_li_k} \\
& + D_i \x_{r_{s-k-l}u_li_{k-1}} - 2r\,\g_{ii} \x_{r_{s-k-l}u_{l+1}i_{k-2}} + 2r\, \g_{ii} \x_{r_{s-k-l+1}u_{l}i_{k-2}} \, ,
\end{split}
\ee
must satisfy (for $n \geq 2$)
\be \label{boundary-summary}
\d \vf_{r\m_{s-1}} = 0 \, , \qquad
\d \vf_{u_{s-k}i_k} = \cO(r^{-\frac{n}{2}+k}) \, .
\ee
When $n =1$, one should have instead $\d \vf_{u_s} = \cO(1)$ and $\d \vf_{u_{s-1}\phi} = \cO(1)$, while all other variations must vanish.

From the examples discussed in the previous sections, we are led to consider the ansatz
\be \label{ansatz-par}
\x_{r_{s-k-l-1}u_li_k} = r^{2k+l} \l_{i_k}{}^{\!\!\!(k,l)}(u,x^m) + \cO(r^{2k+l-1}) \, .
\ee
Under these conditions, the terms in the second line of {the variations} \eqref{variation-s} become subleading for $l>0$, while~the first line gives a first-order differential equation in $u$ which fixes the $u$-dependence of the $\l^{(k,l)}$. 
To~proceed, one can notice that the gauge parameters generating the residual symmetry must be both divergenceless and traceless. These constraints are indeed necessary to leave the gauge-fixed version~\eqref{maxwell-like-s} of the equations of motion invariant. They imply
\be \label{divergence}
\begin{split}
\nabla\cdot \x_{r_{s-k-l-1}u_{l-1}i_k} & = -\, \dot{\x}_{r_{s-k-l}u_{l-1}i_k} - \frac{1}{r} \left( r\pr_r + n \right) \x_{r_{s-k-l-1}u_li_k} + \frac{1}{r^2}\, \Ddot \x_{r_{s-k-l-1}u_{l-1}i_k} \\
& - \frac{s-k-l-1}{r^3}\, \x^{\,\pe}{}_{\!\!\!r_{s-k-l-2}u_{l-1}i_k} + \frac{1}{r} \left( r\pr_r + n \right) \x_{r_{s-k-l}u_{l-1}i_k} = 0 \, ,
\end{split}
\ee
and 
\be \label{trace}
\x_{r_{s-k-l}u_{l-1}i_k} - 2\, \x_{r_{s-k-l-1}u_li_k} + \frac{1}{r^2}\, \x^{\,\pe}{}_{\!\!\!r_{s-k-l-2}u_{l-1}i_k} = 0 \, ,
\ee
where we denoted with a prime a contraction with the $n$-dimensional metric $\g_{ij}$.
Combining this information with the requirement that {the variations} \eqref{variation-s} vanish at leading order, one obtains
\be \label{fixing-u}
\dot{\l}_{i_k}{}^{\!\!\!(k,l)} + \frac{s-k-l-1}{n+s+k-2}\, \Ddot \l_{i_k}{}^{\!\!\!(k+1,l)} = 0 \, ,
\ee
which, as anticipated, fixes the $u$-dependence of the leading order in the expansion in powers of $r$.

 From~now on, we focus on the components of the gauge parameter that are relevant to the computation of asymptotic charges. In Appendix~\ref{app:charges}, we shall show that, in the Bondi gauge \eqref{Bondi-s}, they can be expressed in terms of the non-vanishing field components as
\be \label{charges-s} 
\begin{split}
Q(u) & = -\,\lim_{r\to\infty} \int_{S_u} \frac{r^{n-1} d\Omega_n}{(s-1)!}\, \sum_{p\,=\,0}^{s-1} \bin{s-1}{p} \bigg\{ \vf_{u_{s-p}i_{p}}\left( r\pr_r + n + 2p \right) \x^{u_{s-p-1}i_p} \\
& +  \x^{u_{s-p-1}i_p} \left[ (s-p-2)  \left( r\pr_r + n \right) \vf_{u_{s-p}i_{p}} - \frac{s-p-1}{r}\, D\cdot \vf_{u_{s-p-1}i_p} \right] \bigg\} \, .
\end{split}
\ee

Therefore, they only depend on the components $\x_{r_{s-k-1}i_k} = (-1)^{s-k-1} \x^{u_{s-k-1}}{}_{i_k}$ of the gauge parameters of asymptotic symmetries. According to \eqref{ansatz-par} and \eqref{fixing-u}, these satisfy
\be \label{K-u}
\x_{r_{s-k-1}i_k} = r^{2k} \left( K_{i_k}{}^{\!\!\!(k)} + \sum_{m\,=\,1}^{s-k-1}  \frac{(-1)^m u^m}{(n+s+k-2)_m} \bin{s-k-1}{m} (\Ddot)^m K_{i_k}{}^{\!\!\!(k+m)} \right) + \cO(r^{2k-1}) \, ,
\ee
where the tensors $K^{(k)}(x^m)$ only depend on the coordinates on the $n$-dimensional sphere at null infinity, while $(a)_n \equiv a (a+1) \cdots (a+n-1)$ is the Pochhammer symbol. Moreover, at the leading order in $r$, \eqref{trace} implies that the $K^{(k)}$ are traceless (with respect to contractions with $\g_{ij}$). This implies that, in three dimensions, only $K^{(0)}$ and $K^{(1)}$ are actually present, in analogy with the corresponding reduction in the number of integration constants.

 The tensors $K^{(k)}$ must also satisfy suitable differential constraints, which generalise those displayed in \eqref{K} and \eqref{R} for the spin-three case. To identify them, it is convenient to focus on the variations of the field components without any $u$ index. The absence of $u$-derivatives indeed allows one to study the resulting equations order by order in an expansion in powers of $u$. We can thus introduce the following expansion of the relevant components of the gauge parameter:\footnote{We already encoded the information on the minimum power of $r$ entering the decomposition that can be extracted starting from the inspection of the equation $\d \vf_{r_{s}} = 0$ and substituting the result in the other variations of the form $\d \vf_{r_{s-k}i_k}$.}
\begin{align}
\x_{r_{s-k-1}i_k} & = \sum_{l\,=\,0}^k r^{2k-l} A_{i_k}{}^{\!\!\!(k,l)}(x^m) + \textrm{terms in $u$} \, , \label{xi-ri} \\
\x_{r_{s-k-2}ui_k} & = \sum_{l\,=\,0}^{k+1} r^{2k-l+1} B_{i_k}{}^{\!\!\!(k,l)}(x^m) + \textrm{terms in $u$} \, , \label{xi-rui}
\end{align}
where $A^{(k,0)} \equiv K^{(k)}$ as dictated by \eqref{K-u}.
Substituting this decomposition in {the variations} \eqref{variation-s} gives
\be \label{var-phi_rik}
\begin{split}
\d \vf_{r_{s-k-1}i_{k+1}} = \sum_{l\,=\,0}^k r^{2k-l} \Big\{ & -(l+1)(s-k-1)\, A_{i_{k+1}}{}^{\!\!\!\!\!\!\!\!\!(k+1,l+1)} + D_i A_{i_k}{}^{\!\!\!(k,l)} \\
& - 2\, \g_{ii} B_{i_{k-1}}{}^{\!\!\!\!\!\!\!\!\!(k-1,l)} + 2\, \g_{ii} A_{i_{k-1}}{}^{\!\!\!\!\!\!\!\!\!(k-1,l-1)} \Big\} + \textrm{terms in $u$} \, .
\end{split}
\ee
Preserving the boundary conditions \eqref{boundary-summary} thus requires, for all $k < s-1$,
\be
A_{i_k}{}^{\!\!\!(k,l)} = \frac{1}{l(s-k)}\, D_i A_{i_{k-1}}{}^{\!\!\!\!\!\!\!\!\!(k-1,l-1)} + \g_{ii} \left( \cdots \right)
\quad \Rightarrow \quad
A_{i_{s-1}}{}^{\!\!\!\!\!\!\!\!\!(s-1,l)} = \frac{1}{(l!)^2}\, (D_i)^l K_{i_{s-l-1}}{}^{\!\!\!\!\!\!\!\!\!\!\!\!\!\!(s-l-1)} + \g_{ii} \left( \cdots \right) .
\ee
Notice that, for any value of $l$, the omitted combination in the $A^{(s-1,l)}$ tensor contains at least one new tensor of the $B^{(k,l)}$ family with respect to those appearing for lower values of $l$. As a result, by~substituting back in \eqref{variation-s} (with $k=s-1$), one obtains the condition
\be \label{var-phi_s}
\d \vf_{i_s} = \sum_{l\,=\,0}^{s-1} \frac{r^{2s-l-2}}{(l!)^2} \left\{ (D_i)^{l+1} K_{i_{s-l-1}}{}^{\!\!\!\!\!\!\!\!\!\!\!\!\!\!(s-l-1)} + \g_{ii}\, \X_{\,i_{s-2}}{}^{\!\!\!\!\!\!\!\!\!(s-l-1)}\right\} + \textrm{terms in $u$} = \cO(r^{-\frac{n}{2}+s}) \, ,
\ee
where the tensors $\X^{(k)}$ can be considered as independent.\footnote{For instance, the dependence of the $\X^{(k)}$ on the $K^{(k)}$ tensors can be eliminated by redefining $B^{(k,l)} \to B^{(k,l)} + A^{(k,l-1)}$ in \eqref{xi-rui}.} 

 If the space-time dimension is greater than four, i.e.,\ if $n>2$, the traceless tensors $K^{(k)}$ defined in \eqref{K-u} must therefore satisfy the differential constraints
\be \label{diff-constr-s}
\underbrace{D_i \cdots D_i}_{s-k} K_{i_{k}}{}^{\!\!\!(k)} + \g_{ii}\, \Xi_{\,i_{s-2}}{}^{\!\!\!\!\!\!\!\!\!(k)} = 0 \, , 
\qquad \g^{mn} K_{i_{k-2}mn}^{(k)} \, , 
\qquad 0 \leq k \leq s-1 \, ,
\ee
which, in particular, imply that the $(s-k)$th trace of $\X^{(k)}$ vanishes. Actually, these tensors can be eliminated by computing successive traces of eqs.~\eqref{diff-constr-s}. For instance, for $k=s-1$, we obtained the conformal Killing tensor equation on the sphere (see e.g.\ \cite{conformal-killing}); one can eliminate the tensor $\X^{(s-1)}$ by computing a trace of \eqref{diff-constr-s} to obtain
\be \label{killing-s}
D_i K_{i_{s-1}}{}^{\!\!\!\!\!\!\!\!\!(s-1)} - \frac{2}{n+2s-4}\, \g_{ii} \Ddot K_{i_{s-2}}{}^{\!\!\!\!\!\!\!\!\!(s-1)} = 0 \, , \qquad
\g^{kl} K_{i_{s-3}kl}^{\,(s-1)} = 0 \, ,
\ee
which is the formulation that we used in \eqref{kill2} and \eqref{K} for fields of spin two and three, respectively.

 When $n = 2$, the last term in the sum \eqref{var-phi_s} does not have to vanish in order to preserve the boundary conditions. As a result, the function on the sphere denoted by $K^{(0)}(x^m)$ is completely arbitrary, as~already pointed out in \cite{super}. This infinite-dimensional enhancement of the asymptotic symmetries, generalising BMS supertranslations, is accompanied by a local infinite-dimensional enhancement of the symmetries generated by the tensors $K^{(s-1)}(x^m)$. 
The conformal Killing tensor equation \eqref{killing-s} admits in general $\frac{1}{s+1}\bin{n+s+1}{n+1}\bin{n+s}{n}$ independent globally defined solutions for $n \geq 2$ \cite{conformal-killing}.
When $n = 2$, in addition, one also finds the further, local solutions
\be
K_{z\cdots z}^{(s-1)} = K(z) \, , \qquad
K_{\bar{z}\cdots \bar{z}}^{(s-1)} = \tilde{K}(\bar{z}) \, , \qquad
K_{z \cdots z\, \bar{z}\cdots \bar{z}}^{(s-1)} = 0 \, ,
\ee
where $z = e^{i x^1} \cot \fr{x^2}{2}$ together with its conjugate provide complex coordinates on the celestial sphere. These local solutions provide an extension to arbitrary values of the spin of superrotations  \cite{Barnich_Revisited,Barnich_BMS/CFT}. Moreover, for $s=3$, in \cite{super}, we have shown that the equation for the tensor $K^{(s-2)}$ also admits locally infinitely many solutions. We defer to future work a complete analysis of eq.~\eqref{diff-constr-s}, but it is tempting to conjecture that all these equations admit infinitely-many solutions when the dimension of space-time is equal to four, that is when $n = 2$.

 When $n = 1$, the equations for the surviving tensors $K^{(0)}(\phi)$ and $K^{(1)}(\phi)$ trivialise. As a result, for~each spin-$s$ field, the asymptotic symmetries are generated by two arbitrary functions of the angular coordinate on the circle at null infinity, in analogy with the spin-three case \cite{spin3-3D_1,spin3-3D_2}. The number of free functions in the asymptotic symmetries is also the same as in $AdS_3$ higher-spin theories \cite{W3,Wextra1,Wextra2,Wextra3,metric3D}.

 Before moving to the actual evaluation of the linearised charges, let us mention that one should complete our preliminary analysis of the asymptotic symmetries by studying the cancellation of the subleading orders in {the gauge transformations} \eqref{variation-s}, as we did for $s=3$ in \cite{super}. Nevertheless, we stress that the resulting (naively overdetermined) system of equations admits at least the solutions corresponding to rank-\mbox{$(s-1)$} traceless and divergenceless Killing tensors of the Minkowski background. They indeed generate spin-$s$ gauge transformations leaving the Minkowski background invariant, so that they obviously  satisfy the weaker conditions \eqref{boundary-summary}. Consequently, in analogy with the spin-three example, besides {the differential constraints} \eqref{diff-constr-s}, we do not expect any additional constraint on the tensors $K^{(k)}$ that fully characterises the asymptotic symmetries.

 We now return to the expression \eqref{charges-s} for the asymptotic charges, aiming to make manifest their $u-$independence for $n \neq 2$. While performing this analysis, one first has to keep track of the cancellation of all potentially divergent terms in the limit $r\to\infty$. As we have seen in the previous sections, these usually occur after integrations by parts; for instance, the coefficient of the leading power in $r$ is of the form
\be
	\int_{S_u} d\Omega_n\: K^{(s-1)}_{i_{s-1}} \Ddot C^{(s,0)\,i_{s-1}} = \int_{S_u} d\Omega_n\:  
	D_i^{\phantom{(s)}}\!\!\!\! 
	K^{(s-1)}_{i_{s-1}} C^{(s,0)\,{i_s}} \, ,
\ee 
and hence vanishes because $K^{(s-1)}$ satisfies the conformal Killing tensor equation \eqref{killing-s} and $C^{(s,0)}$ is traceless.
	Although we do not perform here an exhaustive inspection of these cancellations, the~systematics suggested by the examples of spin one, two and three leads us to expect that they hold in general. Assuming that this is indeed the case, the finite contribution to the charges is determined by the Coulomb-like terms $\cQ^{(k)}$ in the boundary conditions \eqref{boundary-cond-s}. Substituting them into {the charge formula} \eqref{charges-s}, while~taking into account their dependence on the integration constants in \eqref{finalQ}, one obtains after integration by parts
\begin{align}
& Q(u) = \int_{S_u} \!\frac{d\Omega_n}{(s-1)!} \sum_{p=0}^{s-1} \sum_{m=0}^{s-p-1} \sum_{l=0}^{p} \bin{s-1}{p} \bin{s-p-1}{m} \bin{p}{l} \frac{(s+p+n-2)(s+n+p-l-2)!}{(s+n+p-2)_m (s+n+p-2)!} \nn \\
& \phantom{Q(u) = \int_{S_u}} \times (-1)^{s+m+p+l}\, u^{l+m} \cM^{(p-l)}_{i_{p-l}} (\Ddot)^{l+m} K^{(m+p)\,i_{p-l}} \label{charges-int-s} \\[5pt]
& = \int_{S_u} \sum_{k,q} \frac{d\Omega_n\,(-1)^{s+k}(s+n+q-2)!}{(s-1)!(s+n+k+q-3)!} \left[ \sum_{p=0}^{s-1} (-1)^p \bin{s-1}{p} \bin{s-p-1}{k+q-p} \bin{p}{q} \right] u^{k} \cM^{(q)}_{i_{q}} (\Ddot)^{k} K^{(k+q)\,i_{q}} \, . \nn
\end{align}

The final expression has been obtained by introducing the new labels $k = l+m$ and $q = p-l$ and by swapping the sums, whose ranges precisely correspond to the values of the labels for which the integrand does not vanish (with the convention $\bin{N}{n} = 0$ for $n < 0$ or $n > N$). One can eventually verify that the sum within square brackets in the second line of \eqref{charges-int-s} vanishes for any $k > 0$,  thus~providing, via the disappearance of the $u-$dependence, a good consistency check of our formulae for any value of the spin. The $u$-independent contribution then reads
\be \label{charges-final}
Q = \int_{S^n} \frac{d\Omega_n}{(s-1)!} \sum_{q\,=\,0}^{s-1} (-1)^{s+q} (s+n+q-2) \bin{s-1}{q}\, K^{(q)}_{i_q} \cM^{(q)\,i_q} \, ,
\ee
in full analogy with the result that we presented for $s=3$ in \eqref{finalcharge3}.\footnote{In order to compare with the spin-two charge \eqref{final2}, one should take into account the factor of $2$ introduced by the definition of the Bondi mass aspect (see e.g.~\eqref{solB2}) and that $T = - K^{(0)}$ as dictated by \eqref{xi2}.}

 When the dimension of spacetime is equal to three, similar considerations apply, with the additional simplification that the tensors $K^{(k)}$ with $k\geq 2$ and the integration constants $\cM^{(k)}$ with $k \geq 2$ are actually absent. By defining $K^{(0)} \equiv T$, $\cM^{(0)} = \cM$ and $K^{(1)}_\phi \equiv v(\phi)$, $\cM^{(1)}_\phi \equiv \cN(\phi)$, in this case {the charge} \eqref{charges-final} becomes
\be
Q = \frac{(-1)^s}{(s-2)!} \int d\phi\, \Big\{  T(\phi) \cM(\phi) - s\, \r(\phi) \cN(\phi) \Big\}
\ee
for any value of spin. In four dimensions, additional terms in $u$, depending on the data of the radiation solution, do appear. In particular, in agreement with the discussion in sect.~3.2 of \cite{super}, the charge formula \eqref{charges-final} receives the following additional contribution in the supertranslation sector:
\be
Q(u) = \frac{(-1)^s s}{(s-1)!} \int_{S_u}\! d\Omega_n\ K^{(0)} \left( \cM^{(0)} + \frac{1}{s!}\, (\Ddot)^s C^{(0)} \right) + \cdots \, .
\ee
The contributions of the radiation data to the terms involving (generalised) superrotations can be determined following the same steps as in Section~\ref{sec:symm3}.

%%%%%%%%%%%%%%%%%%%%%%%%%%%%%%%%%%%%%%%%%%
\vspace{6pt} 
%%%%%%%%%%%%%%%%%%%%%%%%%%%%%%%%%%%%%%%%%%
\acknowledgments{A.C. acknowledges the support of the Universit\'e libre de Bruxelles, where part of this work has been done. His work has been partially supported by the ERC Advanced Grant High-Spin-Grav, by~FNRS-Belgium (convention FRFC PDR T.1025.14 and convention IISN 4.4503.15) and by the NCCR SwissMAP, funded by the Swiss National Science Foundation. A.C., D.F. and C.H. are grateful to the Service de M\'ecanique et Gravitation,
Universit\'e de Mons and to Kyung Hee University, Seoul for the kind hospitality extended to some or all of us when part of this work was being done.  The work of D.F. and of C.H. has been supported in part by Scuola Normale Superiore and by INFN Pisa.}

\appendix

\vspace{6pt}
%%%%%%%%%%%%%%%%%%%%%%%%%%%%%%%%%%%%%%%%%%%%%%%%%%%%%%%%%%%%%%%%%%%%%
\section{Spin-\boldmath{$s$} Charges in Bondi Gauge}\label{app:charges}
%%%%%%%%%%%%%%%%%%%%%%%%%%%%%%%%%%%%%%%%%%%%%%%%%%%%%%%%%%%%%%%%%%%%%

%\numberwithin{equation}{section}
\renewcommand{\theequation}{A\arabic{equation}}
\setcounter{equation}{0}

In this appendix, we consider the linearised charges at null infinity associated with bosonic gauge fields of arbitrary spin. We work in generic space-time dimension and we express the charges in terms of the non-vanishing components of the fields in the Bondi gauge \eqref{Bondi-s}.

%%%%%%%%%%%%%%%%%%%%%%%%%%%%%%%%%%%%%
\subsection{On-Shell Closed Two-Form for Arbitrary Spin}\label{sec:2-form}
%%%%%%%%%%%%%%%%%%%%%%%%%%%%%%%%%%%%%

We begin from the following on-shell closed two-form, which gives the spin-$s$ linearised charges upon integration on a codimension-two surface \cite{charges_spin-s}:
\be \label{covariant}
\begin{split}
k^{[\a\b]} & = \frac{\sqrt{-g}}{(s-1)!} \bigg\{ \nabla^{[\a} \vf^{\b]}{}_{\m_{s-1}} \x^{\m_{s-1}} + \vf_{\m_{s-1}}{}^{[\a} \nabla^{\b]} \x^{\m_{s-1}} + (s-1)\, \nabla\cdot\vf_{\m_{s-2}}{}^{[\a} \x^{\b]\m_{s-2}} \\
& + (s-1)\, \x^{\m_{s-2}[\a} \nabla^{\b]} \vf_{\m_{s-2}} + \frac{s-1}{2} \left(\, \vf_{\m_{s-2}} \nabla^{[\a} \x^{\b]\m_{s-2}} - \nabla_{\!\m} \vf_{\m_{s-3}}{}^{[\a} \x^{\b]\m_{s-2}} \,\right) \bigg\} \, .
\end{split}
\ee
As in Section~\ref{sec:spin-s}, groups of symmetrised indices have been denoted by a single Greek letter with a~label denoting the total number of indices, while repeated indices denote a symmetrisation involving the minimum number of terms needed and without any overall factor. Furthermore, omitted indices denote a trace, that is $\vf_{\m_{s-2}} = g^{\a\b} \vf_{\a\b\m_{s-2}}$.
Eq.~\eqref{covariant} has been obtained by eliminating the triplet auxiliary fields from eq.~(48) of \cite{charges_spin-s} and it applies to any space-time dimension $D$. If the field satisfies Fronsdal's equations of motion and the gauge parameter satisfies the Killing tensor equation $\nabla_{\!\m} \x_{\m_{s-1}} = 0$, then~$\nabla_{\!\a} k^{[\a\b]} = 0$.

We are interested in the charges at null infinity, which are defined as an integral over the sphere of dimension $n = D - 2$ at each point $u$, which we denote by $S_u$. As a result, in the Bondi coordinates~\eqref{bondi-coord}, they involve only a specific component of {the two-form} \eqref{covariant} and they are defined as
\be \label{charge}
Q(u) = \lim_{r \to \infty} \int_{S_u} k^{ur}[\vf,\x]\, dx^{1} \cdots dx^{n} \, .
\ee
%

%%%%%%%%%%%%%%%%%%%%%%%%%%%%%%%%%%%%%
\subsection{Rewriting in Bondi Gauge} 
%%%%%%%%%%%%%%%%%%%%%%%%%%%%%%%%%%%%%
 
We now wish to manifest the simplifications of \eqref{covariant} that are induced by the conditions \eqref{Bondi-s} defining the Bondi gauge.
First of all, the terms in its second line involve the trace of the field and therefore vanish in Bondi gauge. Since the only non-vanishing Christoffel symbols are those displayed in \eqref{christoffel} and the non-vanishing components of the inverse metric are
\be
g^{ur} = - 1 \, , \qquad
g^{rr} = 1 \, , \qquad
g^{ij} = r^{-2} \g^{ij} \, ,
\ee
these conditions also imply
\be
\nabla_{\!u}\,\vf_{r\m_{s-1}} = \nabla_{\!r}\,\vf_{r\m_{s-1}} = \g^{ij} \nabla_{\!i}\,\vf_{rj\m_{s-2}} = 0 \, .
\ee

As a result, in the ``Bondi gauge'', the relevant component of {the two-form} \eqref{covariant} reads
\be \label{intermezzo}
k^{ur} = \frac{r^n \sqrt{\g}}{(s-1)!} \left\{\, \x^{\r_{s-1}} \nabla_{\!r}\, \vf_{u\r_{s-1}} - \vf_{u\r_{s-1}} \nabla_{\!r}\, \x^{\r_{s-1}} + (s-1)\, \x^{u\r_{s-2}} \nabla\cdot\vf_{u\r_{s-2}} \,\right\} .
\ee 
Using \eqref{christoffel} and \eqref{Bondi-s}, one can see that the covariant derivatives that are relevant for \eqref{intermezzo} are
\begin{subequations}
\begin{align}
\nabla_{\!r} \vf_{u_{s-p}i_{p}} & = \frac{1}{r} \left( r\pr_r - p \right) \vf_{u_{s-p}i_{p}} \, , \\
\nabla_{\!r} \x^{u_{s-p-1}i_{p}} & = \frac{1}{r} \left( r\pr_r + p \right) \x^{u_{s-p-1}i_{p}} \, , \\
\nabla\cdot \vf_{u_{s-p-1}i_{p}} & = -\,\frac{1}{r} \left( r\pr_r + n \right) \vf_{u_{s-p}i_{p}} + \frac{1}{r^2}\, D\cdot \vf_{u_{s-p-1}i_{p}} \, ,
\end{align}
\end{subequations}
where we recall that $D_i$ denotes the Levi--Civita connection for the metric $\g_{ij}$ on the sphere at null~infinity. 
All in all, by expanding \eqref{intermezzo} in components, one eventually gets:
\be \label{final-s}
\begin{split}
k^{ur} & = -\,\frac{r^{n-1} \sqrt{\g}}{(s-1)!}\,\sum_{p=0}^{s-1} \bin{s-1}{p} \Big[\, (s-p-2)\, \x^{u_{s-p-1}i_p} \left( r\pr_r + n \right) \vf_{u_{s-p}i_{p}} \\ 
& + \vf_{u_{s-p}i_{p}}\left( r\pr_r + n + 2p \right) \x^{u_{s-p-1}i_p} - \frac{s-p-1}{r}\, \x^{u_{s-p-1}i_p}\, D\cdot \vf_{u_{s-p-1}i_p}  \,\Big] \, .
\end{split}
\ee 
%
%\newpage

%%%%%%%%%%%%%%%%%%%%%%%%%%%%%%%%%%%%%%%%%%
\bibliographystyle{mdpi}

%=====================================
% References, variant A: internal bibliography
%=====================================
\renewcommand\bibname{References}

%%%%%%%%%%%%%%%%%%%%%%%%%%%%%%%%%%%%%%%%%%
\end{document}